\documentclass[letterpaper,twocolumn,10pt]{article}

\usepackage{cite}
\usepackage{amsmath,amssymb,amsfonts}
\usepackage{algorithmic}
\usepackage{graphicx}
\usepackage{textcomp}
\usepackage{xcolor}
\usepackage{xspace}
\usepackage{makecell}
\usepackage{multirow}
\usepackage{enumitem}
\usepackage[numbers]{natbib}
\PassOptionsToPackage{hyphens}{url}
\usepackage{hyperref}
\usepackage[symbol]{footmisc}
\usepackage{booktabs}
\usepackage{usenix2019_v3_margins_arxiv}

\newcommand{\attack}{\mathsf{Light Commands}}

\newcommand{\parhead}[1]{\vspace{0.5pt plus 2pt minus 1pt}\par\noindent\textbf{#1}\hspace{1em plus 0.5em minus 0.5em}}

\begin{document}

\title{Light Commands: Laser-Based Audio Injection Attacks \\ on Voice-Controllable Systems
}

\makeatletter
\newcommand{\linebreakand}{%
\end{@IEEEauthorhalign}
\hfill\mbox{}\par
\mbox{}\hfill\begin{@IEEEauthorhalign}
}
\makeatother

\author{
	{\rm Takeshi Sugawara}\\
	The University of Electro-Communications\\
\href{mailto:sugawara@uec.ac.jp}{{sugawara@uec.ac.jp}} 
	\and
	{\rm Benjamin Cyr}\\
	University of Michigan\\
\href{mailto:bencyr@umich.edu}{{bencyr@umich.edu}} 
	\and
	{\rm Sara Rampazzi}\\
	University of Michigan\\
\href{mailto:srampazz@umich.edu}{{srampazz@umich.edu}} 
	\and
	{\rm Daniel Genkin}\\
	University of Michigan\\
\href{mailto:genkin@umich.edu}{{genkin@umich.edu}} 
	\and
	{\rm Kevin Fu}\\
	University of Michigan\\
\href{mailto:kevinfu@umich.edu}{{kevinfu@umich.edu}} 
}

\maketitle

\begin{abstract}
We propose a new class of signal injection attacks on microphones by physically converting light to sound. We show how an attacker can inject arbitrary audio signals to a target microphone by aiming an amplitude-modulated light at the microphone's aperture. We then proceed to show how this effect leads to a remote voice-command injection attack on voice-controllable systems. Examining various products that use Amazon's Alexa, Apple's Siri, Facebook's Portal, and Google Assistant, we show how to use light to obtain control over these devices at distances up to 110 meters and from two separate buildings. Next, we show that user authentication on these devices is often lacking, allowing the attacker to use light-injected voice commands to unlock the target's smartlock-protected front doors, open  garage doors, shop on e-commerce websites at the target's expense, or even unlock and start various vehicles connected to the target's Google account (e.g., Tesla and Ford). Finally, we conclude with possible software and hardware defenses against our attacks.
\end{abstract}

\section{Introduction}

The consistent growth in computational power is profoundly changing the way that humans and computers interact.  Moving away from traditional interfaces like keyboards and mice, in recent years computers have become sufficiently powerful to understand and process human speech. Recognizing the potential of quick and natural human-computer interaction, technology giants such as Apple, Google, Facebook, and Amazon have each launched their own large-scale deployment of 
voice-controllable (VC) systems that continuously listen to and act on human voice commands. 

With tens of millions of devices sold with Alexa, Siri, Portal, and Google Assistant, users can now interact with services without the need to sit in front of a computer or type on a mobile phone. Responding to this trend, the Internet of Things (IoT) market has also undergone a small revolution. Rather than having each device be controlled via a dedicated manufacture-provided software, IoT manufacturers can  now spend their time making hardware, coupling it with a lightweight interface to integrate their products with Alexa, Siri or Google Assistant. Thus, users can receive information and control products by the mere act of speaking without the need for physical interaction with keyboards, mice, touchscreens, or even buttons.

However, while much attention is being given to improving the capabilities of VC systems, much less is known about the resilience of these systems to software and hardware attacks. Indeed, previous works~\cite{diao2014your, jang2014a11y} already highlight the lack of proper user authentication as a major limitation of voice-only interaction,  causing systems to execute commands from potentially malicious sources.

While early command-injection techniques were noticeable by the device's legitimate owner, more recent works~\cite{carlini2016hidden, DBLP:conf/ccs/ZhangYJZZX17,roy2017backdoor, roy2018inaudible, yuan2018commandersong,vaidya2015cocaine,cisse2017houdini, song2017inaudible} focus on stealthy injection, preventing users from hearing or recognizing the injected commands.

The absence of voice authentication has resulted in a proximity-based threat model, where close-proximity users are considered legitimate, while attackers are kept at bay by physical obstructions like walls, locked doors, or closed windows. For attackers aiming to surreptitiously gain control over physically-inaccessible systems, existing injection techniques are unfortunately limited, as the current state of the art~\cite{roy2018inaudible} has a range of about 25 \textit{ft} (7.62 m) in open space, with physical barriers (e.g., windows) further reducing the distance. 
Thus, in this paper we tackle the following questions:

\medskip
{\centering\emph{
Can commands be remotely and stealthily injected into a voice-controllable system from large distances? If so, how can an attacker perform such an attack under realistic conditions and with limited physical access? Finally, what are the implications of such command injections on third-party IoT hardware integrated with the voice-controllable system?
}}
 \begin{figure*}[t]
	\begin{center}
		\includegraphics[width=1\linewidth]{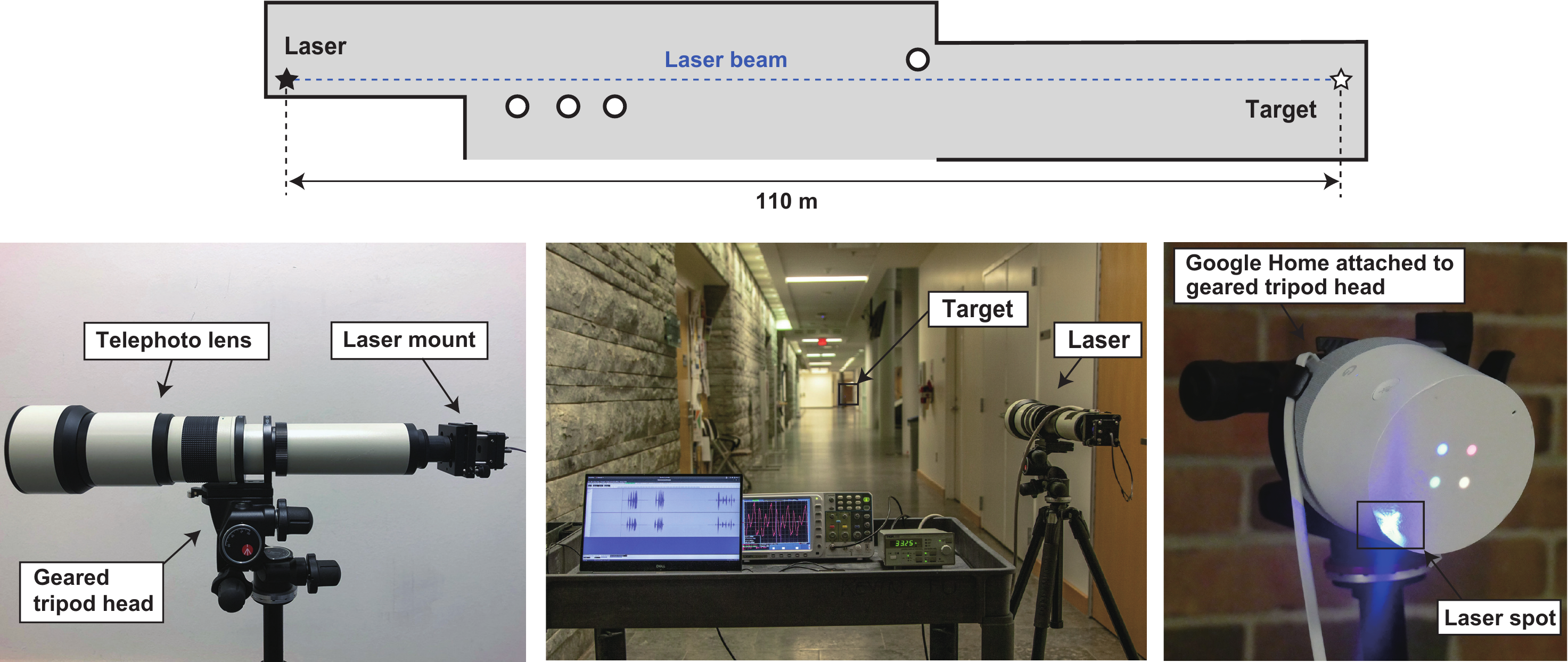}
	\end{center}
	\vspace{-1.9em}
	\caption{Experimental setup for exploring attack range. (Top) Floor plan of the 110 m long corridor. (Left) Laser with telephoto lens mounted on geared tripod head for aiming. (Center) Laser aiming at the target across the 110 m corridor. (Right) Laser spot on the target device mounted on tripod.}
	\label{fig:corridor}
\end{figure*}

\subsection{Our Contribution}
In this paper we present  $\attack$, an attack that can covertly inject commands into voice-controllable systems at long distances. 
\parhead{Laser-Based Audio Injection.} First, we have identified a semantic gap between the physics and specifications of microphones, where microphones often unintentionally respond to light as if it was sound. Exploiting this effect, we can inject sound into microphones by simply modulating the amplitude of a laser light.

\parhead{Attacking Voice-Controllable Systems.} Next, we investigate the vulnerability of popular VC systems (such as Alexa, Siri, Portal, and Google Assistant) to light-based audio injection attacks. We find that 5 mW of laser power (the equivalent of a laser pointer) is sufficient to control many popular voice-activated smart home devices, while about 60 mW is sufficient for gaining control over phones and tablets.

\parhead{Attack Range.} Using a telephoto lens to focus the laser, we demonstrate the first command injection attack on VC systems which achieves distances of up to 110 meters (the maximum distance safely available to us) as shown in Figure~\ref{fig:corridor}. We also demonstrate how light can be used to control VC systems across buildings and through closed glass windows at similar distances. Finally, we note that unlike previous works that have limited range due to the use of sound for signal injection, the range obtained by light-based injection is only limited by the attacker's power budget, optics, and aiming capabilities.

\parhead{Insufficient Authentication.}  Having established the feasibility of malicious control over VC systems at large distances, we investigate the security implications of such attacks. We find that VC systems often lack any user authentication mechanisms, or if the mechanisms are present, they are incorrectly implemented (e.g., allowing for PIN brute forcing). We show how an attacker can use light-injected voice commands to unlock the target's smart-lock protected front door, open garage doors, shop on e-commerce websites, or even locate, unlock and start various vehicles (e.g., Tesla and Ford) if the vehicles are connected to the target's Google account.

\parhead{Attack Stealthiness and Cheap Setup.} We then show how an attacker can build a cheap yet effective injection setup, using commercially available laser pointers and laser drivers.  Moreover, by using infrared lasers and abusing volume features (e.g., whisper mode for Alexa devices) on the target device, we show how an attacker can mount a light-based audio injection attack while minimizing the chance of discovery by the target's legitimate owner.

\parhead{Countermeasures.} Finally, we discuss software and hardware-based countermeasures against our attacks.  

\parhead{Summary of Contributions.} In this paper we make the following contributions.
\begin{enumerate}[leftmargin=*,nolistsep]
	\item Discover a vulnerability in MEMS microphones, making them susceptible to light-based signal injection attacks (Section~\ref{injecting_sound_via_laser_light}).
	\item Characterize the vulnerability of popular Alexa, Siri, Portal, and Google Assistant devices to light-based command injection across large distances and varying laser power (Section~\ref{sec:many-systems}).
	\item Assess the security implications of malicious command injection attacks on VC systems and demonstrate how such attacks can be mounted using cheap and readily available equipment (Section~~\ref{attack_scenarios}).
	\item Discuss software and hardware countermeasures to light-based signal injection attacks (Section~\ref{sect_defense}).
\end{enumerate}

\subsection{Safety and Responsible Disclosure}
\parhead{Laser Safety.}
Laser radiation requires special controls for safety, as high-powered lasers might cause hazards of fire, eye damage, and skin damage. We urge that researchers  receive formal laser safety training and approval of experimental designs before attempting reproduction of our work. In particular, all the experiments in this paper were conducted under a Standard Operating Procedure which was approved by our university's Safety Committee.  

\parhead{Disclosure Process.}
Following the practice of responsible disclosure, we have shared our findings with Google, Amazon, Apple, Facebook, August, Ford, Tesla, and Analog Devices, a major supplier of MEMS microphones. We subsequently maintained contact with the security teams of these vendors, as well as with ICS-CERT and the FDA. The findings presented in this paper were made public on the mutually-agreed date of November 4th, 2019.

\vspace{-1em}
\section{Background} \label{sect_background}

\subsection{Voice-Controllable System}
The term ``Voice-Controllable (VC) system'' refers to a system that is controlled primarily by voice commands directly spoken by users in a natural language, e.g., English.
 While some important exceptions exist, VC systems often immediately operate on voice commands issued by the user without requiring further interaction. For example, when the user commands the VC system to ``open the garage door'', the garage door is immediately opened. 

Following the terminology of \cite{DBLP:conf/ccs/ZhangYJZZX17}, a typical VC system is composed of three main components: (i) voice capture, (ii) speech recognition, and (iii) command
execution. First, the voice capture
subsystem is responsible for converting sound produced by the user into electrical signals. Next, the speech recognition subsystem is responsible for detecting the wake word in the acquired signal (e.g., ``Alexa", ``OK Google", "Hey Portal" or ``Hey Siri") and subsequently interpreting the meaning of the
voice command using signal and natural-language processing. 
Finally, the command-execution subsystem launches the
corresponding application or executes an operation based on the
recognized voice command.

\subsection{Attacks on Voice-Controllable Systems}
Several previous works explored the security of VC systems, uncovering vulnerabilities that allow attackers to issue unauthorized voice commands to these devices~\cite{carlini2016hidden, DBLP:conf/ccs/ZhangYJZZX17,roy2017backdoor, roy2018inaudible, yuan2018commandersong}. 

\parhead{Malicious Command Injection.}
More specifically, \cite{diao2014your,jang2014a11y} developed malicious smartphone applications that play synthetic audio commands into nearby VC systems without requiring any special operating system permissions. While these attacks transmit commands that are easily noticeable to a human listener, other works~\cite{carlini2016hidden,vaidya2015cocaine,cisse2017houdini} focused on camouflaging commands in audible  signals, attempting to make them unintelligible or unnoticeable to human listeners, while still being recognizable to speech recognition models.

\parhead{Inaudible Voice Commands.}
A more recent line of work focuses on completely hiding the voice commands from human listeners.   \citet{roy2017backdoor} demonstrate that high frequency sounds inaudible to humans can be recorded by commodity microphones. Subsequently, \citet{song2017inaudible} and \textit{DolphinAttack}~\cite{DBLP:conf/ccs/ZhangYJZZX17}  extended the work of~\cite{roy2017backdoor}  by sending inaudible commands to VC systems 
 via word modulation on ultrasound carriers. 
 By exploiting microphone nonlinearities, a signal modulated onto an ultrasonic carrier is demodulated to the audible range by the targeted microphone, recovering the original voice command while remaining undetected by humans.

 However, both attacks are limited to short distances
(from 2 \textit{cm} to 175 \textit{cm}) due to the transmitter operating at low power. Unfortunately, 
increasing the transmitting power generates an audible frequency component containing the (hidden) voice command, as the 
transmitter is also affected by the same nonlinearity observed in the receiving microphone. 
Tackling the distance limitation, \citet{roy2018inaudible} 
 mitigated this effect by splitting the signal in multiple frequency bins and playing them through an array of 61 speakers. However, the re-appearance of audible leakage still limits the attack's range to 25 \textit{ft} (7.62 m) in open space, with physical barriers (e.g., windows) and the absorption of ultrasonic waves in air further reducing range by attenuating the transmitted signal.

\parhead{Skill Squatting Attacks.}
A final line of work focuses on confusing speech recognition systems, causing them to misinterpret correctly-issued voice commands. 
These so-called skill squatting attacks~\cite{kumar2018skill,zhang2018understanding} work by exploiting systematic errors in the recognition of similarly sounding words, routing users to malicious applications without their knowledge.

\subsection{Acoustic Signal Injection Attacks}
Several works used acoustic signal injection as a method of inducing unintended behavior in various systems.

More specifically, \citet{son2015rocking} showed that MEMS sensors are sensitive to ultrasound signals, resulting in denial of service attacks against inertial measurement unit (IMU) on drones. Subsequently, \citet{yan2016can} demonstrated that acoustic waves can be used to saturate and spoof ultrasonic sensors, impairing car safety. This was further improved by Walnut~\cite{DBLP:conf/eurosp/TrippelWXHF17}, which exploited aliasing and clipping effects in the sensor's components to achieve precise control over MEMS accelerometers via sound injection. 

More recently, \citet{DBLP:conf/ccs/NashimotoSSS18} showed the possibility of using sound to attack sensor-fusion algorithms that rely on data from multiple sensors (e.g., accelerometers, gyroscopes, and magnetometers) while Blue Note~\cite{DBLP:conf/sp/BoltonRLKXF18} demonstrates the feasibility of sound attacks on mechanical hard drives, resulting in operating system crashes.

\subsection{Laser Injection Attacks} \label{sect_laser_injection_attacks}
In addition to sound, light has also been utilized for signal injection. 
Indeed,  \cite{petit2015remote,6899663,yan2016can} mounted denial of service attacks on 
 cameras and LiDARs by illuminating victims' photo-receivers with strong lights. This was later extended by \citet{DBLP:conf/ches/ShinKKK17} and \citet{cao2019adversarial} to a more sophisticated attack that injects precisely-controlled
 signals to LiDAR systems, causing the target to see an illusory object. Next, \citet{park2016ain} showed an attack on medical infusion pumps, using light to attack optical sensors that count the number of administered medication drops. Finally, \citet{uluagac2014sensory} show how various sensors, such as infrared and light sensors, can be used to activate and transfer malware between infected devices.

Another line of work focuses on using light for injecting faults inside computing devices, resulting in security breaches. More specifically, it is well-known that laser light causes soft (temporary) errors in semiconductors, where similar errors are also caused by ionizing radiation~\cite{Habing65}.  Exploiting this effect, \citet{DBLP:conf/ches/SkorobogatovA02} showed the first light-induced fault attacks on smartcards and microcontrollers, 
demonstrating the possibility of flipping individual bits in  memory cells. This effect was subsequently exploited in numerous follow ups, using laser-induced faults to compromise the hardware's data and logic flow, extract  secret keys, and dump the device's memory. See \cite{KSV13,dutertre2011review} for further details.

\begin{figure*}[t]
	\begin{center}
		\includegraphics[width=1\linewidth]{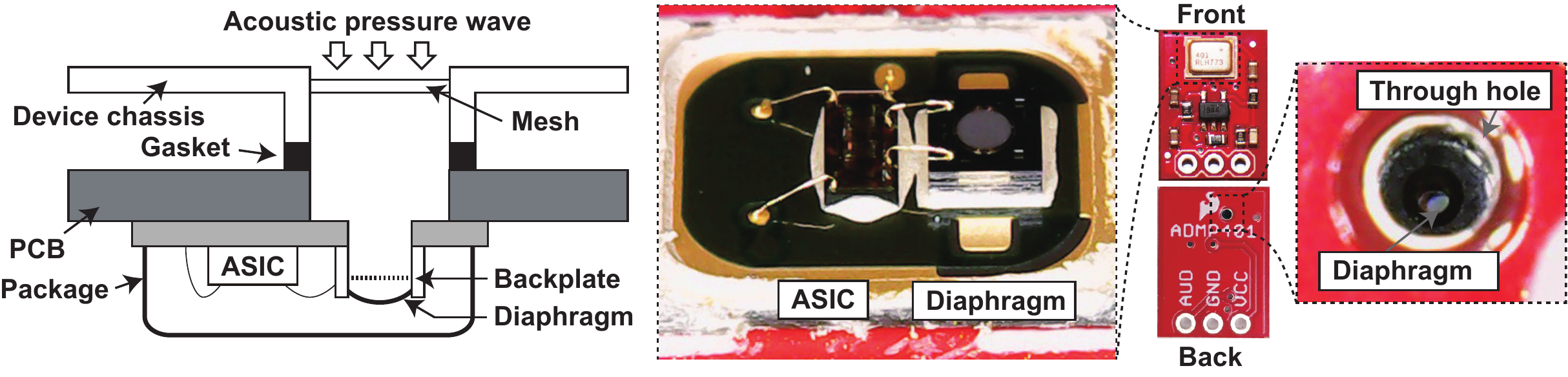}
	\end{center}
\vspace{-1.99em}
	\caption{MEMS microphone construction. (Left) Cross-sectional view of a MEMS microphone on a device. (Middle) A diaphragm and ASIC on a depackaged microphone. (Right) Magnified view of an acoustic port on PCB.
} \label{fig:microphone}

\end{figure*}

\subsection{MEMS Microphones}
MEMS is an integrated implementation of mechanical components on a
 chip, typically fabricated with an etching process. While there are a number of different MEMS sensors (e.g., accelerometers and gyroscopes), in this paper we focus on MEMS-based microphones, which are particularly popular in mobile and embedded applications (such as smartphones and smart speakers) due to their small footprints and low prices. 

\parhead{Microphone Overview.} The left column of Figure~\ref{fig:microphone} shows the construction of a typical backport MEMS microphone, which is composed of a diaphragm and an ASIC circuit. The diaphragm is a thin membrane that flexes in response to an acoustic wave. The diaphragm and a fixed back plate work as a parallel-plate capacitor, whose capacitance changes as a consequence of the diaphragm's mechanical deformations as it responds to alternating sound pressures. Finally, the ASIC die converts the capacitive change to a voltage signal on the output of the microphone.

\parhead{Microphone Mounting.} A backport MEMS microphone is mounted on the surface of a printed circuit board (PCB), with the microphone's aperture exposed through a cavity on the PCB (see the third column of Figure~\ref{fig:microphone}). The cavity, in turn, is part of an acoustic path that guides sound through holes (acoustic ports) in the device's chassis to the microphone's aperture. 
Finally, the device's acoustic ports typically have a fine mesh as shown in Figure~\ref{fig:acoustic_port} to prevent dirt and foreign objects from entering the microphone.

\begin{figure}[t]
	\begin{center}
		\includegraphics[width=1\linewidth]{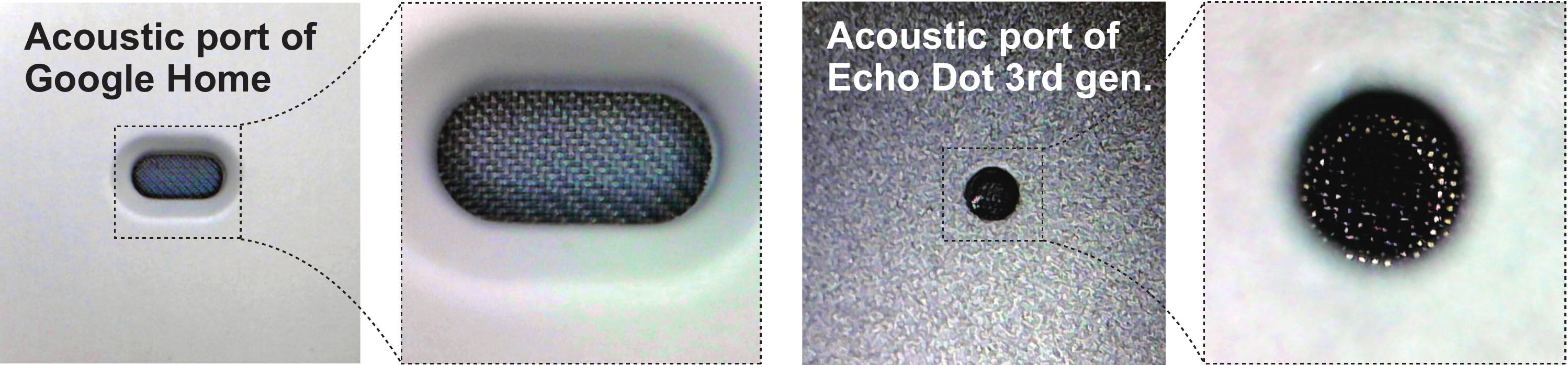}
	\end{center}
\vspace{-1.99em}
	\caption{Acoustic port of (Left) Google Home and (Right) Echo Dot 3rd generation. The ports are located on the top of the devices, and there are meshes inside the port.}
 \label{fig:acoustic_port}
 \vspace{-1em}
\end{figure}

\subsection{Laser Sources}\label{sect_laser_sources}

\parhead{Choice of a Laser.}
A laser is a device that emits a beam of coherent light that stays narrow over a long distance and be focused to a tight spot. While other alternatives exist, in this paper we focus on laser emitting diodes, which are common in consumer laser products such as laser pointers. Next,  as the light intensity emitted from a laser diode is directly proportional to the diode's driving current, we can easily encode analog signals via the beam's intensity by using a laser driver capable of amplitude modulation.

	\parhead{Laser Safety and Availability.}
As strong, tightly focused lights can be potentially hazardous, there are standards in place regulating lights emitted from laser systems~\cite{iec,FDA_standard} that divide lasers into classes based on the potential for injury resulting from beam exposure.  In this paper, we are interested in two main types of devices, which we now describe. 

\parhead{Low-Power Class 3R Systems.} This class contains devices whose output power is less than 5 mW at visible wavelength (400--700 nm, see Figure~\ref{fig:spectrum}). While prolonged intentional eye exposure to the beam emitted from these devices might be harmful, these lasers are considered safe for brief eye-exposures. As such, class 3R systems form a good compromise between safety and usability, making these lasers common in consumer products such as laser pointers. 

\parhead{High-Power Class 3B and Class 4 Systems.} Next, lasers that emit between 5 and 500 mW are classified as class 3B systems, and might cause eye injury even from  short beam exposure durations. 
Finally, lasers that emit over 500 mW of power are categorized as class 4, which can instantaneously cause blindness, skin burns and fires. As such, uncontrolled exposure to class 4 laser beams should be strongly avoided. 

However, despite the regulation, there are reports of high-power class 3B and 4 systems being openly sold as ``laser pointers''~\cite{menace}. While purchasing laser pointers from Amazon and eBay, we have discovered a troubling discrepancy between the rated and actual power of laser products. While the labels and descriptions of most products stated an output power of 5 mW, the actual measured power was sometimes as high as 1 W (i.e., $\times 200$ above the allowable limit).

\begin{figure}[t]
	\begin{center}
		\includegraphics[width=\linewidth]{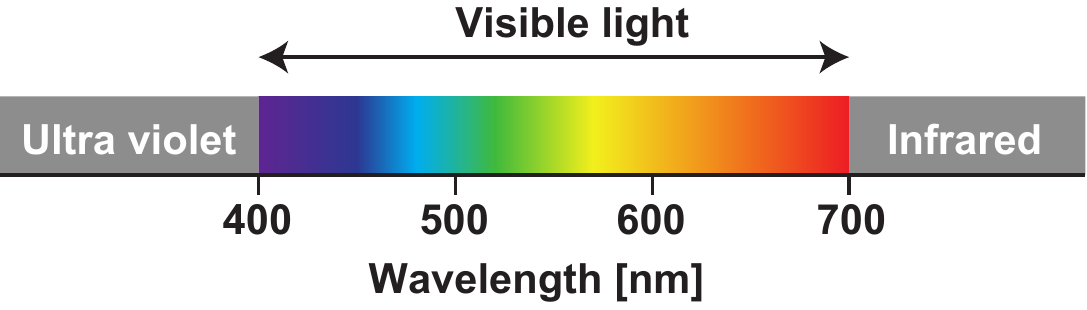}
	\end{center}
\vspace{-1.9em}
	\caption{Wavelength and color of light}
 \label{fig:spectrum}
\end{figure}

\section{Threat Model} \label{threat_model}
The attacker's goal is to remotely inject malicious commands into the targeted voice-controllable device without being detected by the device's owner. More specifically, we consider the following threat model. 

\parhead{No Physical Access or Owner Interaction.} We assume that the attacker does not have any physical access to the victim device. Thus, the attacker cannot press any buttons, alter voice-inaccessible settings, or compromise the device's software. Finally, we assume that the attacker cannot make the device's owner perform any useful interaction (such as pressing a button or unlocking the screen).

\parhead{Line of Sight.} We do assume however that the attacker has remote line of sight access to the target device and its microphones. We argue that such an assumption is reasonable, as voice-activated devices (such as smart speakers, thermostats, security cameras, or even phones) are often left visible to the attacker, including through closed glass windows. 

\parhead{Device Feedback.} We note that the remote line of sight access to the target device usually allows the attacker to observe the device's LED lights. 
Generally, these LEDs light up after a device properly recognizes its wake-up word (e.g., Alexa, Hey Google) and show unique colors and light patterns once a voice command has been recognized and accepted. Observing the lights, an attacker can use this feedback to remotely determine if an attack attempt was successful.

\parhead{Device Characteristics.}
Finally, we also assume that the attacker has access to a device of a similar model as the target device. Thus, the attacker knows all the target's physical characteristics, such as location of the microphone ports and physical structure of the device's sound path. Such knowledge can easily be acquired by purchasing and analyzing a device of the same model before launching attacks. We do not, however, assume that the attacker has prior access the specific device instance used by the victim. In particular, all the experiments done in this paper were empirically verified to be applicable to other devices of the same model available to us without instance-specific calibration.

\section{Injecting Sound via Laser Light} \label{injecting_sound_via_laser_light}
\subsection{Signal Injection Feasibility}\label{sec:feasibility}
In this section we explore the feasibility of injecting acoustic signals into microphones using laser light. We begin by describing our experimental setup. 

\parhead{Setup.}  
We used a blue Osram PLT5 450B 450-nm laser diode connected to a Thorlabs LDC205C laser driver. We increased the diode's DC current with the driver until it emitted a continuous 5~mW laser beam, while measuring light intensity using the Thorlabs S121C photo-diode power sensor. The beam was subsequently directed to the acoustic port on the SparkFun MEMS microphone breakout board mounting the Analog Devices ADMP401 MEMS microphone. Finally, we recorded the diode current and the microphone's output using a Tektronix MSO5204 oscilloscope, see Figure~\ref{fig:feasibility_experiment}. The experiments were conducted in a regular office environment, with typical ambient noise from human speech, computers, and air conditioning systems. 

\begin{figure*}[t]
	\begin{center}
		\includegraphics[width=\linewidth]{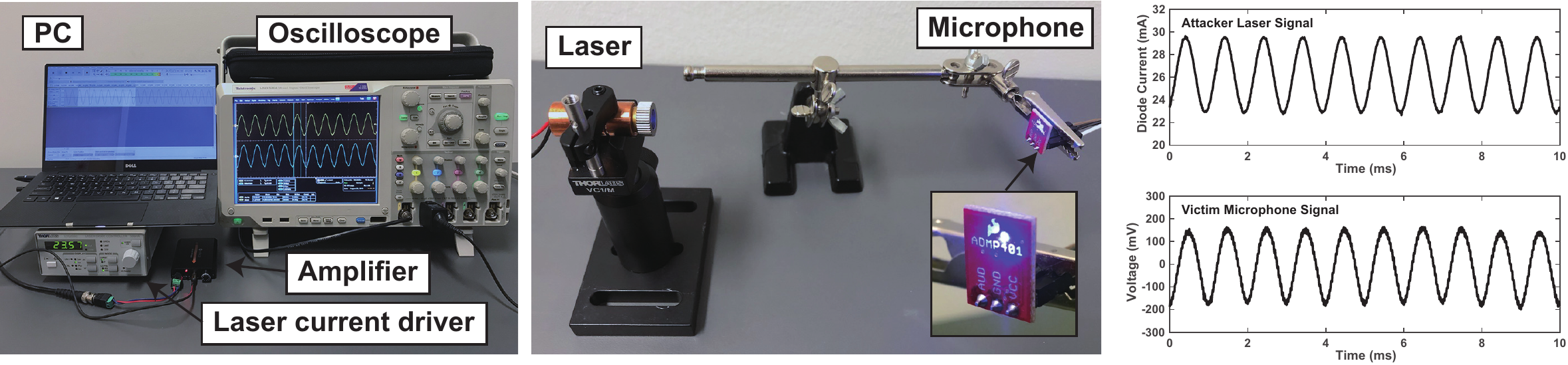}
	\end{center}
\vspace{-1.99em}
	\caption{Testing signal injection feasibility. (Left) A setup for signal injection feasibility composed of a laser current driver, PC, audio amplifier, and oscilloscope. (Middle) Laser diode with beam aimed at
a MEMS microphone breakout board. (Right) Diode current and microphone output waveforms. } \label{fig:feasibility_experiment}
\end{figure*}

\parhead{Signal Injection by Converting Sound to Light.}
To convert sound signals into light, we encode the intensity of the sound signal as the intensity of the laser beam, where louder sounds make for larger changes in light intensity and weaker sounds correspond to smaller changes. Next, as the intensity of the light beam emitted from the laser diode is direction proportional with the supplied current, we use a laser driver to regulate the laser diode's current as a function of an audio file played into the driver's input port. This resulted in the audio waveform being directly encoded in the intensity of the light emitted by the laser.

More specifically,  we used the current driver to modulate a sine wave on top of the diode's current $I_t$ via amplitude modulation (AM), given by the following equation:
\begin{align}
I_t = I_{DC} + \frac{I_{pp}}{2} \sin(2 \pi f t) \label{eq:laser_current-new}
\end{align}
where $I_{DC}$ is a DC bias, $I_{pp}$ is the peak-to-peak amplitude, and $f$ is the frequency. In this section, we set  $I_{DC}$ = 26.2 mA, $I_{pp} = 7$ mA and $f$ = 1 kHz. The sine wave was played using a laptop's on-board soundcard, where the speaker output was connected to the modulation input port on the laser driver via a Neoteck NTK059 audio amplifier. 
The laser driver~\cite{thorlabs} performs an amplitude modulation (AM) 
of the sine wave onto its output current without needing additional custom circuits or software. Finally, as the light intensity emitted by the laser diode is directly proportional to the current provided by the laser driver, this resulted in a 1 kHz sine wave directly encoded in the intensity of the light emitted by the laser diode.

\parhead{Observing the Microphone Output.} As can be seen in Figure~\ref{fig:feasibility_experiment}, the microphone output clearly shows a 1 kHz sine wave that matches the frequency of the injected signal without any noticeable distortion.

\subsection{Characterizing Laser Audio Injection} \label{sect_characterization}
Having successfully demonstrated the possibility of injecting audio signals via laser beams, we now proceed to characterize the light intensity response of the diodes (as a function of current) and the frequency response of the microphone to laser-based audio injection. To see the wavelength dependency, we also examine a 638-nm red laser (Ushio HL63603TG) in addition to the blue one used in the previous experiment.

\parhead{Laser Current to Light Characteristics.} 
We begin by examining the relationship between the diode current and the optical power of the laser. For this purpose, we aimed a laser beam at our Thorlabs S121C power sensor while driving the diodes with DC currents, i.e., $I_{pp}=0$ in Equation~\ref{eq:laser_current-new}. Considering the different properties of the diodes, the blue and red laser are examined up to 300 and 200 mA, respectively. 

The first column of Figure~\ref{fig:laser_char-new} shows the current vs. light (I-L) curves for the blue and red lasers. The horizontal axis is the diode current $I_{DC}$ and the vertical axis is the optical power. As can be seen, once the current provided to the laser is above the diode-specific threshold (denoted by $I_{th}$), the light power emitted by the laser increases linearly with the provided current. Thus, as $|\sin(2\pi ft)| <1$, we have an (approximately) linear conversion of current to light provided that  $I_{DC} - I_{pp}/2 > I_{th}$.

\begin{figure*}[t]
	\begin{center}
		\includegraphics[width=\linewidth]{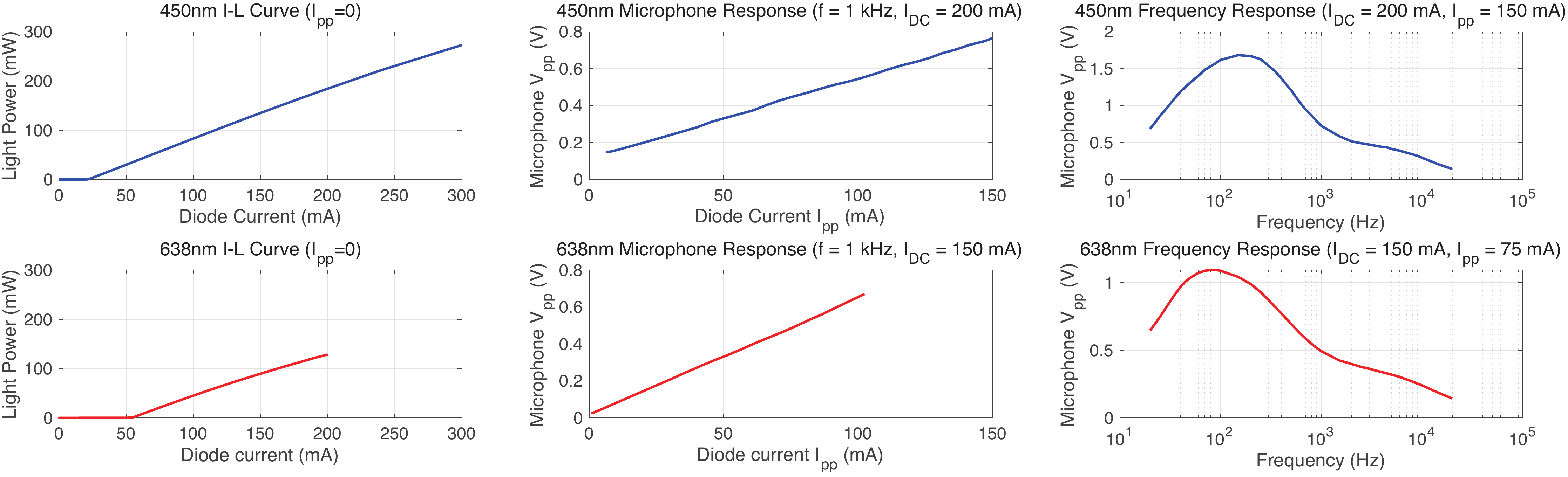}
	\end{center}
\vspace{-1.99em}
	\caption{Characteristics of the 450-nm blue laser (first row) and the 638-nm red laser (second row). 
		(First column) Current-light DC characteristics. (Second column) Microphone response for a 1 kHz tone with different amplitudes. 
(Third column) Frequency responses of the overall setup for fixed bias and amplitude.
} \label{fig:laser_char-new}
\end{figure*}

\parhead{Laser Current to Sound Characteristics.}
We now proceed to characterize the effect of light injection on a MEMS microphone. We achieve this by aiming an amplitude-modulated (AM) laser beam with variable current amplitudes ($I_{pp}$) and a constant current offset ($I_{DC}$) into the
aperture of the Analog Devices ADMP401 microphone, mounted on a breakout board. We subsequently monitor the peak-to-peak voltage of the microphone's output, plotting the resulting signal. 

The second column of Figure~\ref{fig:laser_char-new} shows the relationship between the modulating signal $I_{pp}$ and the resulting signal $V_{pp}$ for both the blue and red laser diodes. 
The results suggest that the driving alternating current $I_{pp}$ (cf. the bias current) is the key for strong injection: we can linearly increase the sound volume received by the microphone by increasing the driving AC current $I_{pp}$.

\parhead{Choosing $I_{DC}$ and $I_{pp}$.} 
Given a laser diode that can emit a maximum average power of $L$ mW, we would like to choose the values for $I_{DC}$ and $I_{pp}$ which result in the strongest possible microphone output signals, while having the average optical power emitted by the laser be less than or equal to~$L$ mW. From the leftmost column of Figure~\ref{fig:laser_char-new}, we deduce that the laser's output power is linearly proportional to the laser's driving current $I_t = I_{DC}+ I_{pp} \sin(2 \pi f t)$, and the average power depends mostly on $I_{DC}$, as $I_{pp} \sin(2 \pi f t)$ averages out to zero. 

Thus, to stay within the power budget of $L$ mW while obtaining the strongest possible signal at the microphone output, the attacker must first determine the DC current offset $I_{DC}$ that results in the diode outputting light at $L$ mW, and then subsequently maximize the amplitude of the microphone's output signal by setting  $I_{pp}/2 = I_{DC} - I_{th}$.\footnote{We note here that the subtraction of $I_{th}$ is designed to ensure that $I_{DC} - I_{pp}/2 > I_{th}$, meaning that the diode stays in its linear region thereby avoiding signal distortion.}

\parhead{Characterizing the Frequency Response of Laser Audio Injection.} Next, we set out to characterize the response of the microphone to different frequencies of laser-injected sound signals. We use the same operating points as the previous experiment, and set the tone's amplitude such that it fits with the linear region ($I_{DC}= 200$~mA and $I_{pp} = 150$~mA for the blue laser, and $I_{DC}= 150$~mA and $I_{pp} = 75$~mA for the red laser). We then record the microphone's output levels while changing the frequency $f$ of the light-modulated sine wave.

The third column of Figure~\ref{fig:laser_char-new} shows the obtained frequency
response for both blue and red lasers. The horizontal axis is the frequency while the vertical axis is the peak-to-peak voltage of the microphone output. Both lasers have very similar responses, covering the entire audible band 20 Hz--20 kHz, implying the possibility of injecting any audio signal. 

\parhead{Choice of Laser.}
Finally, we note the color insensitivity of injection. Although blue and red lights are on the other edges on the visible spectrum (see Figure~\ref{fig:spectrum}), the levels of injected audio signal are in the same range and the shapes of the frequency-response curves are also similar. Therefore, color has low priority in choosing a laser compared to other factors for making $\attack$. In this paper, we consistently use the 450-nm blue laser mainly because of (i) better availability of high-power diodes and (ii) the advantage in focusing because of a shorter wavelength.

\subsection{Mechanical or Electrical Transduction?}
In this section we set out to investigate the physical root  cause behind of the microphone's sensitivity to light. We consider both the photoelectric and photoacoustic effects, and try to distinguish between them by selectively illuminating different parts of the microphone's internal structure using lasers. 
\parhead{Photoelectric Effect.}
Traditional laser fault injection attacks on semiconductor chips (as described in~\ref{sect_laser_injection_attacks}) are explained by the photoelectric effect in transistors~\cite{KSV13,dutertre2011review} resulting in irregularities in the device's digital logic. Likewise, MEMS microphones also have ASICs inside their packages, which are used for converting the capacitive changes of the diaphragm into an electrical signal (see Figure~\ref{fig:microphone}). Such ASICs can be externally-illuminated via lasers through the microphone's exposed acoustic port. As strong light hits a semiconductor chip, it induces a photocurrent across a transistor, where the current's strength is proportional to the light intensity~\cite{Habing65}. The analog part of the microphone's ASIC recognizes this photocurrent as a genuine signal from the diaphragm, resulting in the microphone treating light as sound. Confirming this, while not common in smart speakers, we have seen several other microphone vendors covering the ASIC with opaque resin, known in the industry as ``goop''.

\parhead{Photoacoustic Effect.} 
The light sensitivity of microphones can also be attributed to the photoacoustic effect~\cite{MR16}, which converts optical to kinetic energy and induces mechanical vibration at the illuminated material. The effect is well known for more than 100 years since its discovery by Alexander Graham Bell back in 1880~\cite{Bell1880}, which is now used for spectroscopy and bioimaging. Although we have not found any previous work on the photoacoustic effect specific to a MEMS microphone, the effect is universal and available even with ambient water vapor in the air~\cite{Sullenberger:19}.

\parhead{Selective Laser Illumination.}
We can further narrow the root cause of the microphone's light sensitivity, by noticing that the photoelectric effect happens on an ASIC while the photoacoustic effect on a diaphragm. Thus, by selectively illuminating different microphone components using a laser, we attempted to precisely show the physical root cause.  

We achieve this by opening the metal package of the Analog Devices ADMP401 microphone and injecting analog signals into its diaphragm and ASIC components using a focused laser beam (see Figure~\ref{fig:microphone}). 
After using a microscope to focus a 200~$\mu m$ laser spot on the microphone's components, we observed the strongest signal while aiming the laser on the microphone's ASIC, as shown in Figure~\ref{fig:glue_experiment}(left). This direct injection is very efficient, where less than 0.1~mW of laser power was sufficient to saturate the microphone. We take this as an indication that laser light can cause photoelectric transduction inside the microphone's ASIC, since in our attack the light is reflected onto the ASIC from the microphone's metal package. 
After covering the microphone's ASIC with opaque epoxy (Figure~\ref{fig:glue_experiment}(right)), aiming the laser on the ASIC no longer generates any signal. However, even after the treatment, the microphone still generates a signal when the laser spot is aimed at the microphone's diaphragm. 

Based on these results, we conclude that in addition to the photoelectric effect observed on the microphone's ASIC, there is another light-induced transduction within the MEMS diaphragm. Since the diaphragm is a simple capacitor, we hypothesize that this effect is due to the physical movements of the microphone's diaphragm (i.e., light-induced mechanical vibration).

Next, while the above is not a comprehensive survey on different MEMS microphones, this analysis does provide an overall understanding of the root cause of the physicals effects observed in this paper. Finally, for the experiments conducted in the remainder of this paper, we have aimed the laser through the microphone's acoustic port. We hypothesize that our attacks illuminated both the microphone's ASIC and diaphragm, resulting in some combination of the photoacoustic and photoelectric effects.

\begin{figure}[t]
	\vspace{-1em}
	\begin{center}
		\includegraphics[width=\linewidth]{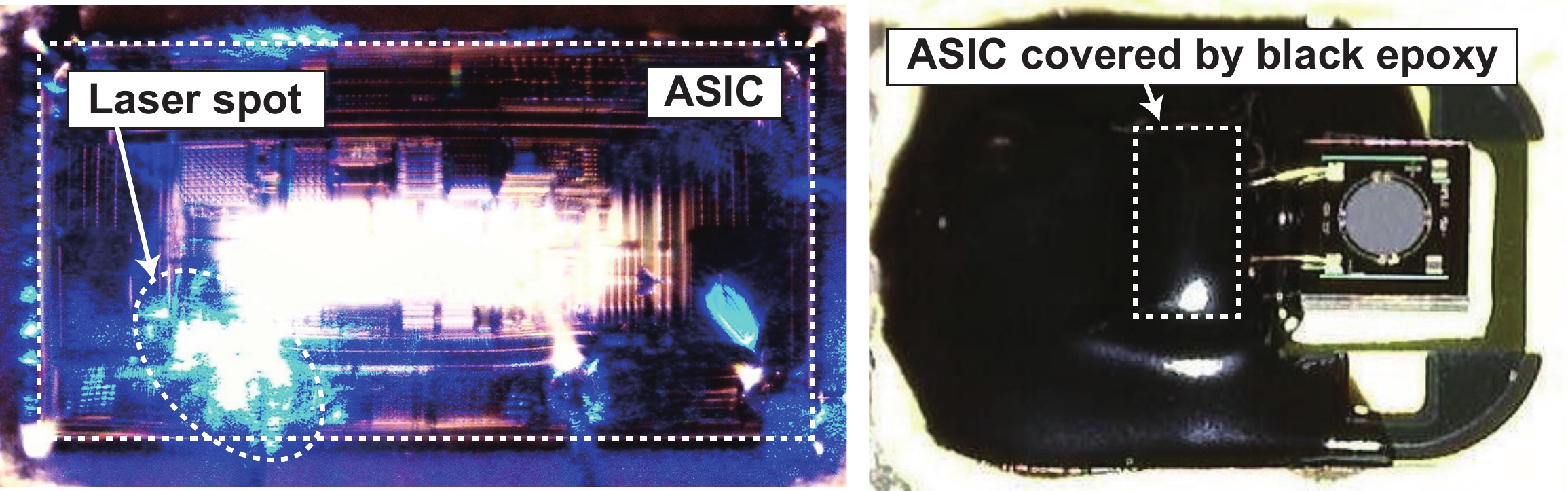}
	\end{center}
\vspace{-1.99em}
	\caption{(Left) Laser spot on the ADMP401's ASIC. (Right) the ASIC covered with opaque epoxy to block laser light.}
	\label{fig:glue_experiment}

\end{figure}

\section{Attacking Voice-Controllable Systems} \label{sec:many-systems}
In this section we evaluate our attack on seventeen popular VC systems. We aim to find out the minimum laser power required by the attacker to gain control over the VC system under ideal conditions as well as the maximum distance that such control can be obtained under more realistic conditions.

\parhead{Target Selection.} We benchmark our attack against several consumer devices which have voice control capabilities (see Table~\ref{tab:distance}). We aim to test the most popular voice assistants -- namely Alexa, Siri, Portal, and Google Assistant.  
While we do not claim that our list is exhaustive, we do argue that it does provide some intuition about the vulnerability of popular VC systems to laser-based voice injection attacks. Next, to explore how different hardware variations (rather than algorithmic variations) affect our attack performance, we benchmark our attack on multiple devices running the same voice recognition backend: Alexa, Siri, Portal and Google Assistant, as summarized in Table~\ref{tab:distance}. For some devices, we examine different generations to explore the differences on attack performance for various hardware models. Finally, we also considered third-party devices with built-in speech recognition, such as the EcoBee thermostat.

\subsection{Exploring Laser Power Requirements} \label{sect_many_speakers}
In this section we aim to characterize the minimum laser power required by the attacker under ideal conditions to control a voice-activated system. Before describing our experimental setup, we discuss our selection of voice commands and experiment success criteria.

\parhead{Command Selection.} We have selected four different voice commands that represent common operations performed by voice-controllable systems.
\begin{itemize}[leftmargin=*, nolistsep]
	\item \parhead{What Time Is It?} We use this command as a baseline of our experiments, as it only requires the device to correctly recognize the command and access the Internet to recover the current time.
	
	\item \parhead{Set the Volume to Zero.} Here, we demonstrate the attacker's ability to control the output of a VC system. We expect this to be the first voice command issued by the attacker, in an attempt to avoid attracting attention from the target's legitimate owner.

	\item \parhead{Purchase a Laser Pointer.} With this command we show how an attacker can potentially place order for various products on behalf  (and at the expense) of users. The attacker can subsequently wait for delivery near the target's residents and collect the purchased item.
	\item \parhead{Open the Garage Door.} Finally, we show how an attacker can interact with additional systems which have been linked by the user to the targeted VC system. While the garage door opener is one such example with clear security implications, we discuss other examples in Section \ref{attack_scenarios}.
\end{itemize}

\parhead{Command Generation.} We have generated audio recordings of all four of the above commands using a common audio recording system (e.g., Audacity). Each command recording was subsequently appended to a recording of the wake word corresponding to the device being tested (e.g., Alexa, Hey Siri, Hey Portal, or OK, Google) and normalized to adjust the overall volume of the recordings to a constant value. We obtained a resulting corpus of 16 complete commands. Next, for each device, we injected four of the complete commands (those beginning with the device-appropriate wake word) into the device's microphone using the setup described below and observed the device's response.
Finally, we note that no device-specific calibration was done during the generation of the audio files containing the voice commands. 

\parhead{Verifying Successful Injection.} We consider a command injection successful in case the device indicates the correct interpretation of the command. We note that 
some commands require other devices to be attached to the victim account in order to properly execute, resulting in an error otherwise (e.g., a garage door opener for a command opening the garage door). 
As in this section we only test feasibility of command injection (as opposed to end-to-end attacks of Section~\ref{attack_scenarios}), we consider an injection attempt successful in case the device properly recognized all the command's words. For devices with screens (e.g., phones and screen-enabled speakers), we verified that the device displayed a correct transcription of the light-injected command. 
Finally, for screen-less devices (e.g., smart speakers), we examined the command log of the account associated with the device for the command transcription.

\parhead{Attack Success Criteria.}
For a given power budget, distance, and command, we consider the injection successful when the device correctly recognized the command during three consecutive attempts. The injection attempt is considered to be a failure otherwise (e.g., the device only recognizes the wake-up word but not the entire command). We take this as an indication that the power budget is sufficient for achieving a near-perfect consecutive command recognition assuming suitable aiming and focusing.

Next, we consider an attack successful, for a given power budget and distance, when all four commands are successfully injected to the device in three consecutive injection attempts. The attack is considered a failure in any other case (e.g., achieving two out of three correct command recognitions). Like in the individual command case, we take this as an indication that the considered power budget and distance is sufficient for a successful command injection. As such, the results in this section should be seen as a conservative estimate of what an attacker can achieve for each device assuming good environmental conditions (e.g., quiet surroundings and suitable aiming).

\parhead{Voice Customization and Security Settings.} For the experiments conducted in this section, we left all the device's settings in their default configuration. In embedded Alexa and Google VC systems (e.g., smart speakers, cameras, etc.) voice customization is off by default,  meaning that the device will operate on commands spoken by any voice. 
Meanwhile, for phones and tablets, we left the voice identification in its default activated setting. For such devices, to ascertain the minimum required power for a successful attack, we personalized the device's voice recognition system with the human voice used to generate the command recordings described above. We then subsequently inject the audio recording of the commands using the same voice without any other customization. Finally, in Section \ref{voice}, we discuss bypassing various voice matching mechanisms.

\parhead{Experimental Setup.} We use the same blue laser and Thorlabs laser driver as in Section~\ref{sec:feasibility}, aiming the laser beam at microphone ports of the devices listed in Table~\ref{tab:distance} from a distance of about 30 cm. 
 Next, to control the surrounding environment, the entire setup was placed in a metal enclosure, with  opaque bottom and sides and with a dark red semi-transparent acrylic top plate, designed to block blue light. See Figure~\ref{fig:setup_speakers}. As the goal of the experiments described in this section is to ascertain the minimum required power for a successful attack on each device, we have used a pair of electrically controlled scanning mirrors (40 Kbps high-speed laser scanning system for laser shows) to precisely place the laser beam in the center of the device's microphone port. Before each experiment we manually focused the laser so that the laser spot size hitting the microphone is minimal. 

For aiming at devices whose microphone port is covered with cloth (e.g., Google Home Mini shown in Figure~\ref{fig:googlehomemini}), the position of the microphone ports can be determined using an easily-observable reference point such as the device's wire connector or LED array. Finally, we note that the distance between the microphone and the reference point is easily obtainable by the attacker either by exploring  his own device, or by referring to online teardown videos~\cite{mini_teardown}.

\begin{figure}[t]
	\begin{center}
		\includegraphics[width=0.99\linewidth]{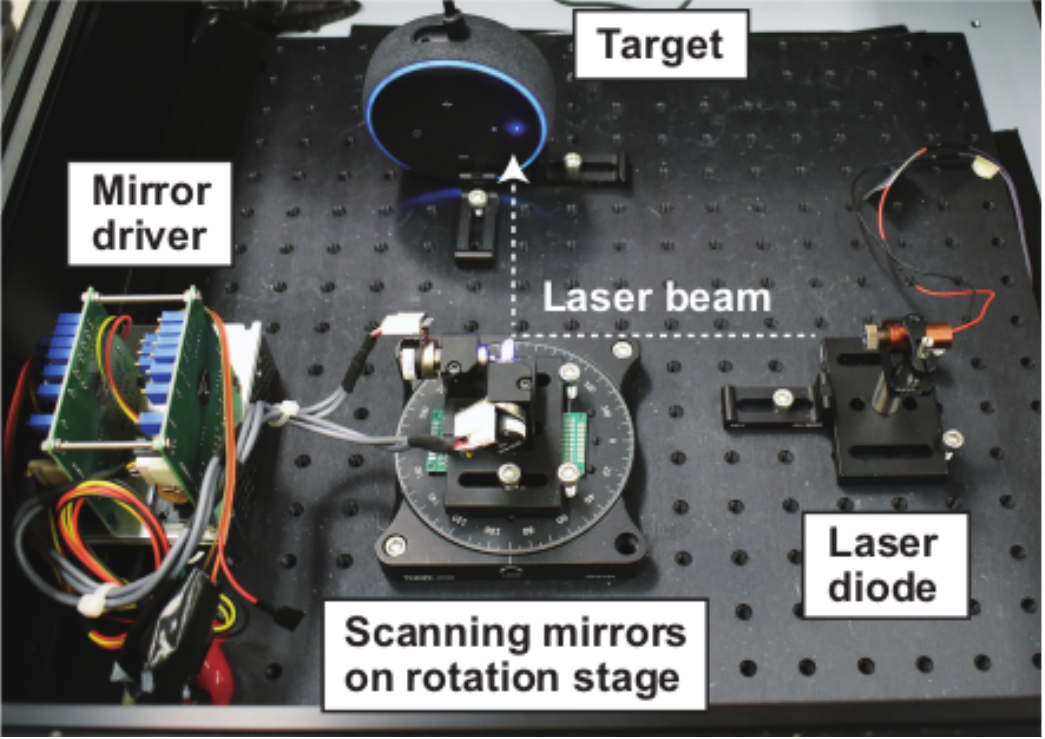}
	\end{center}
\vspace{-1.9em}
	\caption{Exploring minimum laser power requirements: the laser and target are arranged inside an enclosure. The laser spot is aimed at the target acoustic port using electrically controllable scanning mirrors inside the enclosure. The enclosure's top red acrylic cover was removed for visual clarity.} \label{fig:setup_speakers}

\end{figure}

\begin{figure}[t]
	\begin{center}
		\includegraphics[width=0.55\linewidth]{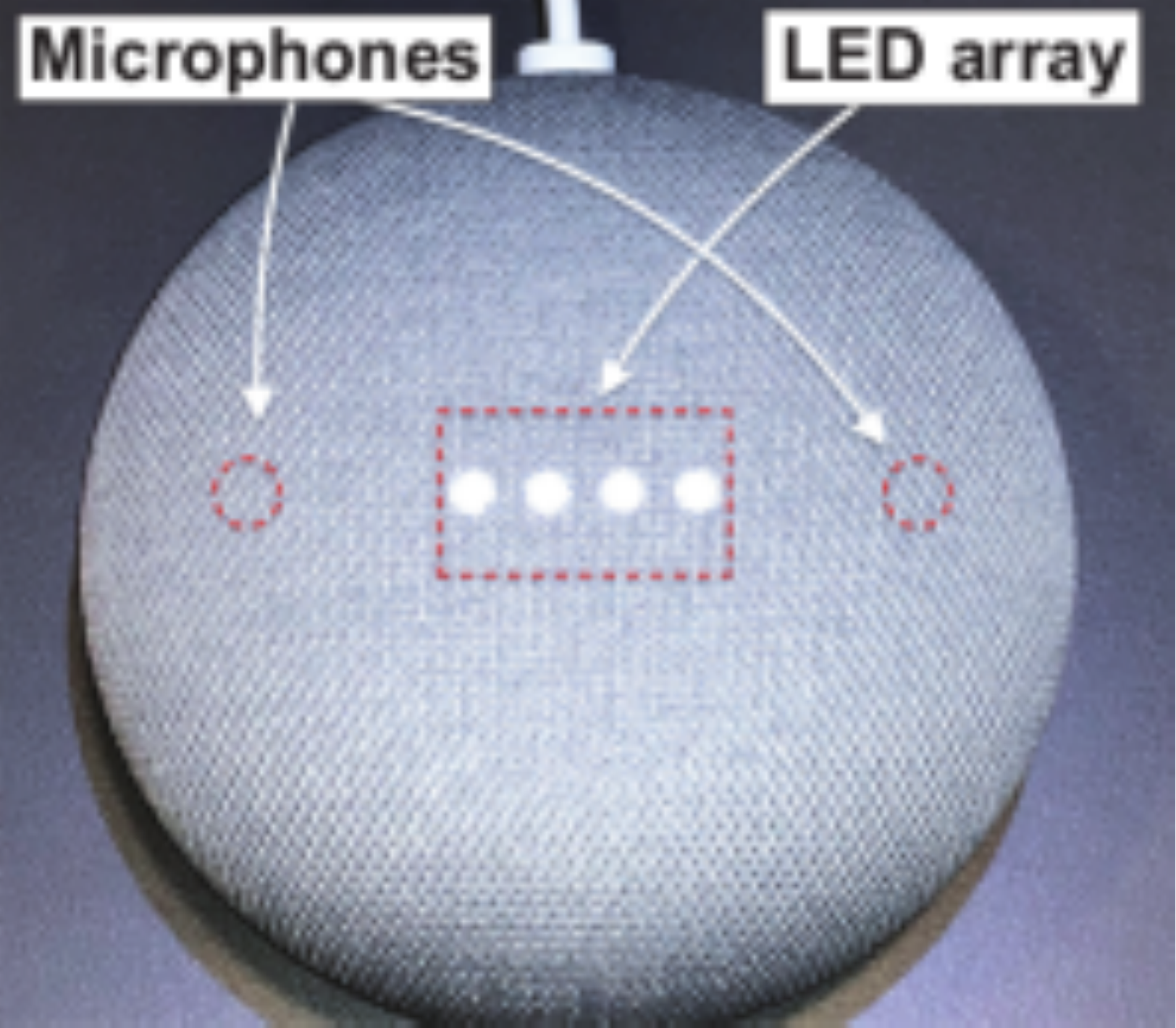}
	\end{center}
\vspace{-1.9em}
	\caption{Google Home Mini. Notice the cloth-covered microphone ports.}
 \label{fig:googlehomemini}
\end{figure}

\begin{table*}[t]
	\caption{Tested devices with minimum activation power and maximum distance achievable at the given power of 5 mW and 60 mW. A 110 m long hallway was used for 5 mW tests while a 50 m long hallway was used for tests at 60 mW.}
	\vspace{-1.5em}
	
	\begin{center}
		\resizebox{\linewidth}{!}{
			\begin{tabular}{lcccccc}
				\toprule
				\multirow{2}{*}{\bf{Device}} & \multirow{2}{*}{\bf{Backend}} & \multirow{2}{*}{\bf{Category}} & \bf{Authen-} & \bf{Minimum} & \bf{Max Distance} &  \bf{Max Distance} \\  
				& & & \bf{tication} & \bf{Power [mW]*} & \bf{at 60 mW [m]**} & \bf{at 5 mW [m]***}  \\
				\midrule
				Google Home                       & Google Assistant & Speaker    &  No &  0.5 &  50+ & 110+ \\ 
				Google Home Mini                  & Google Assistant & Speaker    &  No &  {16} &  20 &  --- \\ 
				Google Nest Cam IQ                & Google Assistant & Camera     &  No &  9 &  50+ &  --- \\ \hline
				Echo Plus 1st Generation          & Alexa            & Speaker    &  No &  2.4 &  50+ & 110+ \\ 
				Echo Plus 2nd Generation          & Alexa            & Speaker    &  No &  2.9 &  50+ & 50 \\ 
				Echo                              & Alexa            & Speaker    &  No &  25 &  50+ & --- \\ 
				
				Echo Dot 2nd Generation           & Alexa            & Speaker    &  No &  7 &  50+ & --- \\ 
				Echo Dot 3rd Generation           & Alexa            & Speaker    &  No &  9 &  50+ &  --- \\ 
				Echo Show 5                       & Alexa            & Speaker    &  No &  17 &  50+ &  --- \\ 
				Echo Spot                         & Alexa            & Speaker    &  No &  29 &  50+ &  --- \\ 
				Facebook Portal	Mini (Front Mic)			  & Alexa 		     & Speaker    &  No & 1 &   50+ & 40 \\ 
				Facebook Portal Mini (Front Mic)$^\S$			  & Portal 			 & Speaker    &  No & 6 &   40 & --- \\ 
				Fire Cube TV                      & Alexa            & Streamer   &  No &  13 &  {20} & --- \\ 
				EcoBee 4                          & Alexa            & Thermostat &  No &  1.7 &  50+ & 70 \\ \hline
				iPhone XR (Front Mic)             & Siri             & Phone      & Yes &  21 &  10 & --- \\ 
				iPad 6th Gen                      & Siri             & Tablet     & Yes & 27 &  20 & --- \\
				Samsung Galaxy S9 (Bottom Mic)    & Google Assistant & Phone      & Yes & 60 &   5 & --- \\ 
				Google Pixel 2 (Bottom Mic) & Google Assistant       & Phone      & Yes & 46 &   5 & --- \\ 
				\bottomrule
				\multicolumn{7}{l}{*at 30 cm distance, **Data limited to a 50 m long corridor, ***Data limited to a 110 m  long corridor, $^\S$Data generated using only the first 3 commands.}
			\end{tabular}
		}
	\end{center}
	\label{tab:distance}
	\vspace{-1em}
\end{table*}

\parhead{Experimental Results.} 
The fifth column of Table~\ref{tab:distance} presents a summary of our results. While the power required from the attacker varies from 0.5 mW (Google Home)  to 60 mW (Galaxy S9), all the devices are susceptible to laser-based command injection, even when the device's microphone port (e.g., Google Home Mini) is covered with fabric and / or foam.

 Finally, for Facebook's Portal Mini device which supports both Amazon's and Facebook's voice assistants, we note the $\times6$ increase in minimum power between ``Hey Portal" and ``Alexa" wakeup words. In addition, Portal also consistently failed to identify the word ``laser'' used in the last command, forcing us to disregard it. As both experiments were done using the same setup and with the laser aimed at the same microphone, we attribute these to algorithmic differences between Amazon's and Facebook's voice recognition backends.

\subsection{Exploring Attack Range} \label{sect_corridor}
The  experiments done in Section~\ref{sect_many_speakers} are performed under ideal conditions, at close range and with the aid of electronic aiming mirrors. Thus, in this section we report on attack results under more realistic distance and aiming conditions.

\parhead{Experimental Setup.}
From the experiments performed in Section~\ref{sect_many_speakers} we note that about 60 mW of laser power is sufficient for successfully attacking all of our tested devices (at least under ideal conditions). Thus, in this section we benchmark the range of our attack using two power budgets.

\begin{itemize}[leftmargin=*,nolistsep]
\item \parhead{60 mW High-Power Laser.} 
As explained in Section~\ref{sect_laser_sources}, we frequently encountered laser pointers whose measured power output was above 60 mW, which greatly exceeds legal 5 mW restrictions. Thus, emulating an attacker which does not follow laser safety protocols for consumer devices,  we benchmark our attack using a 60 mW laser, which is sufficient for successfully attacking all of our tested devices in the previous experiment.

\item \parhead{5 mW Low-Power Laser.}  Next, we also explore the maximum range of a more restricted attacker, which is limited to the maximum amount of power allowed in the U.S. for consumer laser pointers, namely 5~mW.

\end{itemize}

\parhead{Laser Focusing and Aiming.} For large attack distances (tens of meters), laser focusing requires a large diameter lens and cannot be done via the small lenses that are typically used for laser pointers. Thus, we mounted our laser to an Opteka 650-1300 mm high-definition telephoto lens, with 86 mm diameter (Figure~\ref{fig:corridor}(left)). Finally, to simulate realistic aiming conditions for the attacker, we avoided the use of electronic scanning mirrors (used in Section~\ref{sect_many_speakers}) and mounted the lens and laser on a geared camera head (Manfrotto 410 Junior Geared Tripod Head) and tripod. Laser aiming and focusing was done manually, with the target also mounted on a separate tripod. See Figure~\ref{fig:corridor} for a picture of our setup.

\parhead{Test Locations and Experimental Procedure.}
As eye exposure to a 60 mW laser is potentially dangerous, we blocked off a 50 meter long corridor in our office building and performed the experiments at night. However, due to safety reasons, we were unable to obtain a longer corridor for our high-power tests. For lower-power attacks, we performed the experiments in a 110 meter long corridor connecting two buildings (see Figure~\ref{fig:corridor}(top)). In both cases, we fixed the target at increasing distances and adjusted the optics accordingly to obtain the smallest possible laser spot. We regulated the diode current so that the target is illuminated with 5 or 60 mW respectively. 
Finally, the corridor is illuminated with regular fluorescent lamps at office-level brightness while the ambient acoustic noise was about 46~dB (measured using a General Tools DSM403SD sound level meter).

\parhead{Success Criteria.}
We use the same success criteria as in Section~\ref{sect_many_speakers}, considering the attack successful at a given distance in case the device correctly recognized all commands during three consecutive injection attempts and considering failure otherwise. We take this as an indication of the maximum range achievable by the attack at the considered power budget. Finally, we benchmark our attack's accuracy as a function of distance in Section~\ref{sec:benchmarking}. 

\parhead{Experimental Results.} Table~\ref{tab:distance} contains a summary of our distance-benchmarking results. With 60 mW laser power, we have successfully injected voice commands to all the tested devices from a distance of several meters. For devices that can be attacked using 5~mW, we also conducted the low-power experiment in the 110~m hallway. Untested devices are marked by '---' in Table~\ref{tab:distance} due of their high minimum activation power. 

While most devices require a 60 mW laser for successful command injection (e.g., a non-standard-compliant laser pointer), some popular smart speakers such as Google Home and Eco Plus 1st and 2nd Generation are particularly sensitive, allowing for command injection even with 5 mW power over tens of meters. 
Next, as our attacks were conducted in 50 and 110 meter hallways (for 60 and 5 mW lasers, respectively) for some devices, we had to stop the attack when the maximum hallway length was reached. We mark this case with a `+' sign near the device's range in the appropriate column. 

\parhead{Attack Transferability.}
Despite inevitable manufacturing variability between the 17 devices tested in this work, we did not observe any significant changes between the response of different microphones to laser injection. That is, all microphones had shown the same high-level behavior, reacting to light as if it was sound without any microphone-specific calibration. This evidence also supports the universality of our attack, as once the laser was aimed and focused, all devices responded to injected commands without the need for per-device calibration. 
Finally, we note that all devices tested in this paper have multiple microphones, while we aimed our laser to only a single microphone port. However, despite this, the attack is still successful indicating that, at the time of writing, VC systems do not require the microphones' signals to match before executing voice commands.

\subsection{Exploring Attack's Success Probability}\label{sec:benchmarking}
In the attacks presented in Sections~\ref{sect_many_speakers},~\ref{sect_corridor}, and Table~\ref{tab:distance}, all the tested devices properly recognized the injected commands once suitable aiming and focusing were achieved. However, as can be seen in  Table~\ref{tab:distance}, some devices stopped recognizing the commands after exceeding a certain distance. 
Investigating this phenomenon, we explored the attack's error rate
at the borderline attack range. To achieve this, we use a Google Home Mini device as a case study, as its attack range is limited to 20 meters which is shorter than the 50 meter {corridor} available to us for high-power 60~mW experiments. 

Table~\ref{tab:benchmarking} presents a summary of our findings, where each command was injected into the Google Home Mini device 10 times (totaling  40 consecutive command injections). As can be seen, at 20 meters injection attacks are nearly always successful, with a single error in recognizing the word ``laser'' in the third command. However, at 25 meters the success probability significantly falls, with no successful injections observed at 27 meters. These results indicate that while some commands are a slightly harder to inject than others, the sudden drop in 
performance at 27m indicates that our attack's success probability does not seem to be dominated by the command's phonemes. Instead, it appears that success probability is governed by command-unrelated factors such as the internal microphone structure, the presence of fabric covering the microphone ports, the power density of the light hitting the device's microphone ports, the laser beam focus, alignment, environmental noise level, machine learning algorithms,  etc. We leave the task of investigating these factors to future work. 

\begin{table}[t]\centering\small
	\vspace{-1em}
\caption{Attack success accuracy as a function of distance.\label{tab:benchmarking}}
	\begin{tabular}{lccc} 
		\toprule
		\multicolumn{1}{c}{\textbf{Command}} & \textbf{20m} & \multicolumn{1}{c}{\textbf{25m}} & \textbf{27m}  \\ \hline
		{What Time Is It?}          & 100\%        & 90\%                             & 0\%          \\
		{Set the Volume to Zero}           & 100\%        & 80\%                             & 0\%          \\
		{Purchase a Laser Pointer}          & 90\%         & 0\%                              & 0\%         \\
		{Open the Garage Door}          & 100\%        & 100\%                            & 0\%          \\
		\bottomrule
	\end{tabular}

\end{table}

\subsection{Attacking Speaker Authentication} 
\label{voice}
We begin by distinguishing between speaker recognition features, which are designed to recognize voice of specific users and personalize the device's content, and speaker authentication features which is designed to restrict access control to specific users. While not the main topic of this work, in this section we now discuss both features in the context of  light-based command injection.

\parhead{No Speaker Authentication for Smart Speakers.}
We observe that for smart-speaker devices (which are the main focus of this work), speaker recognition is disabled by default at the time of writing. Next, even if the feature is enabled by careful users, smart speakers are designed to be used by multiple users. Thus, their speaker recognition features are usually limited to content personalization rather than authentication, treating unknown voices as guests. Empirically verifying this, we found that Google Home and Alexa smart speakers block voice purchasing for unrecognized voices (presumably as they do not know which account should be billed for the purchase)  while allowing previously-unheard voices to execute security critical voice commands such as unlocking doors.
Finally, we note that at the time of writing, voice authentication (as opposed to personalization) is not available for smart speakers, which are common home smart assistant deployments.

\parhead{Phone and Tablet Devices.}
Next, while not the main focus of this work, we also investigated the feasibility of light command injection into phones and tablets. For such devices, speaker authentication is enabled by default due to the high processing power and single owner use.
	
\parhead{Overview of Voice Authentication.}
After being personalized with samples of the owner's voice speaking specific sentences, the tablet or phone continuously listens to the microphone and acquires a set of voice samples. The collected audio is then used by the device's proprietary voice recognition systems, aiming to recognize the device's owner speaking assistant-specific wake up words (e.g., ``Hey Siri'' or ``OK Google'').
Finally, when there is a successful match with the owner's voice, the phone or tablet device proceeds to execute the voice command.

\parhead{Bypassing Voice Authentication.} Intuitively, an attacker can defeat the speaker authentication feature using  authentic voice recordings of the device's legitimate owner speaking the desired voice commands. Alternatively, if no such recordings are available, DolphinAttack~\cite{DBLP:conf/ccs/ZhangYJZZX17} suggests using speech synthesis techniques, such as splicing relevant phonemes from other recordings of the owner's voice, to construct the commands.  

\parhead{Wake-Only Security.}
However, during our experiments we found that speaker recognition is used by Google and Apple to only verify the wake word, as opposed to the entire command. For example, Android and iOS phones trained to recognize a female voice, correctly execute commands where only the wake word was spoken by the female voice, while the rest of the command was spoken using a male voice. Thus, to bypass voice authentication, an attacker only needs a recording of the 
device's wake word in the owner's voice (which can be obtained by recording any command spoken by the owner).

\parhead{Reproducing Wake Words.} Finally, we explore the possibility of using Text-To-Speech (TTS) techniques for reproducing the owner's voice saying the wake words for a tablet or phone based voice assistant. To that aim, we repeat the phone and tablet experiments done in Sections~\ref{sect_many_speakers},~\ref{sect_corridor} and Table~\ref{tab:distance}, training all the phone and tablet devices with a human female voice. We then used NaturalReader~\cite{NaturalReader}, an online TTS tool for generating the wake words specific for each device, hoping that the features of one of the offered voices will mistakenly match the human voice used for personalization.  See Table~\ref{tab:voice-bruteforcing} for device-specific voice configurations, as provided by NaturalReader, which mistakenly match the female voice used for training. Next, we concatenate the synthetically-generated wake word spoken in a female voice to a voice command pronounced by a male native-English speaker. 
  Using these recordings, we successfully replicated the minimum power and maximum distance results as presented in Table~\ref{tab:distance}. 
  
  We thus conclude that while voice recognition is able to enforce some similarity between the attacker's and owner's voices, it does not offer sufficient entropy to form an adequate countermeasure to command injection attacks. In particular, out of the 18 English voices supported by NaturalReader, we were able to find an artificial voice matching the human female voice used for personalization for all four of our tablets and phones 
 without using any additional customization.  Finally, we did not test the ability to match voices for devices other than phones and tablets, as voice authentication is not available for smart speakers at the time of writing.

\begin{table}
	\vspace{-1em}
	\begin{center}
			        \caption{Bypassing voice authentication on phones and tablets\label{tab:voice-bruteforcing}}
			        \vspace{-1em}
			        \resizebox{\columnwidth}{!}{
\begin{tabular}{llll}
	\toprule
	\multicolumn{1}{l}{\textbf{Device}} & \textbf{Assistant} & \multicolumn{1}{c}{\textbf{TTS Service}} &
	\textbf{Voice Name}  
	\\ \hline
	iPhone XR           & Siri          & NaturalReader                   & US English Heather      \\
 	iPad 6th Gen        & Siri          & NaturalReader                   & US English Laura      \\
	Galaxy S9           & Google Assistant& NaturalReader                 & US English Laura      \\
	Pixel 2           & Google Assistant& NaturalReader                 & US English Laura      \\
	\bottomrule
\end{tabular}
}
	\end{center}
\vspace{-1em}
\end{table}

\section{Exploring Various Attack Scenarios} \label{attack_scenarios}
The results of Section~\ref{sec:many-systems} clearly demonstrate the feasibility of laser-based injection of voice commands into voice-controlled devices across large   attack distances. In this section, we explore the security implications of such an injection, as well as experiment with more realistic attack conditions.

\subsection{A Low-Power Cross-Building Attack}\label{sec:cross-building}
For the long-range attacks presented in Section~\ref{sect_corridor}, we deliberately placed the target device so that the microphone ports are facing directly into the laser beam. While this is realistic for some devices (who have microphone ports on their sides), such an arrangement is artificial for devices with top-facing microphones (unless mounted sideways on the wall). 

In this section we perform the attack under a more realistic conditions, where an attacker aims from another higher building at a target device placed upright on a window sill.

\parhead{Experimental Conditions.} We use the laser diode, telephoto lens and laser driver from Section~\ref{sec:many-systems}, operating the diode at 5~mW (equivalent to a laser pointer) with the same modulation parameters as in the previous section. Next, we placed a Google Home device (which only has top-facing microphones) upright near a window, on a fourth-floor office (15 meters above the ground). The attacker's laser was placed on a platform inside a nearby bell tower, located 43 meters above ground level. Overall, the distance between the attacker's and laser was 75 meters, see Figure~\ref{fig:tower} for the configuration.

\begin{figure*}[t]
	\begin{center}
		\hspace{-1.6em}\includegraphics[width=1.03\linewidth]{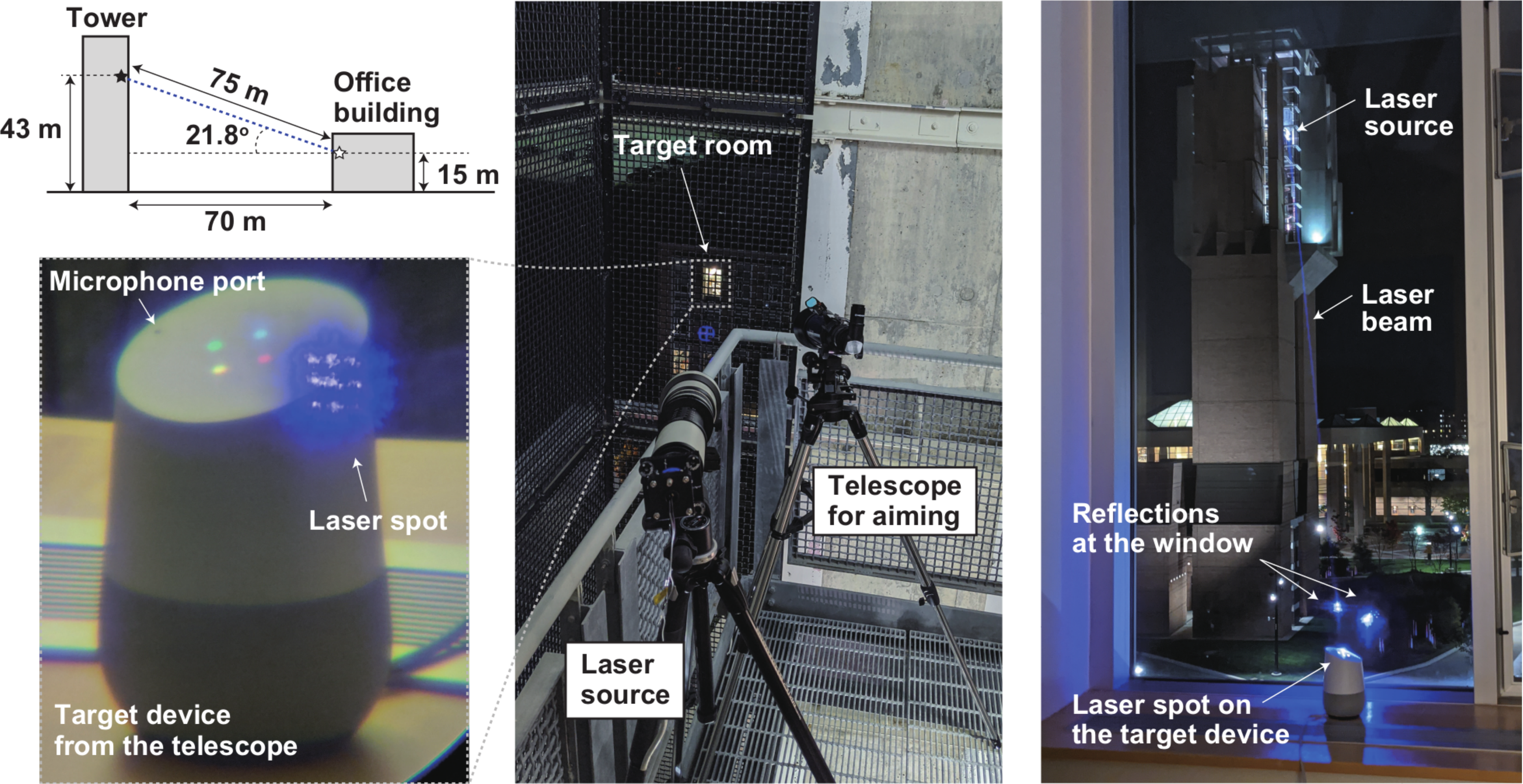}
	\end{center}
	\vspace{-1.99em}
	\caption{Setup for the low-power cross-building attack: (Top left) Laser and target arrangement. (Bottom left) Picture of the target device as visible through the telescope, with the microphone ports and laser spot clearly visible. (Middle) Picture from the tower: laser on telephoto lens aiming down to the target. (Right) Picture from the office building: laser spot on the target device.
	}
	\label{fig:tower}
\end{figure*}

\parhead{Laser Focusing and Aiming.} 
As in Section~\ref{sect_corridor}, it is impossible to focus the laser using the small lens typically used for laser pointers. We thus mounted the laser to an Opteka 650-1300 mm telephoto lens. Next, to aim the laser across large distances, we have mounted the telephoto lens on a Manfrotto 410 geared tripod head. This allows us to precisely aim the laser beam on the target device across large distances, achieving an accuracy far exceeding the one possible with regular (non-geared) tripod heads where the attacker's arm directly moves the laser module. Finally, in order to see the laser spot and the device's microphone ports from far away, we have used a consumer-grade Meade Infinity 102 telescope. As can be seen in Figure~\ref{fig:tower} (left), the Google Home microphone's ports are clearly visible through the telescope.\footnote{Figure~\ref{fig:tower} (left) was taken via a cell phone camera attached to the telescope's eyepiece. Unfortunately, due to imperfect phone-eyepiece alignment, the outcome is slightly out of focus and the laser spot is over saturated. However, the Google Home was in sharp focus with a small laser spot when viewed directly by a human observer.}

\parhead{Attack Results.} We have successfully injected commands into the Google Home target in the above described conditions. We note that despite its low 5~mW power and windy conditions (which caused some beam wobbling due to laser movement), the laser beam successfully injected the voice command while penetrating a closed double-pane glass window. While causing negligible reflections, the double-pane window did not cause any visible distortion in the injected signal, with the laser beam hitting the target's top microphones at an  angle of 21.8 degrees
and successfully injecting the command without the need for any device- or window-specific calibration.
We thus conclude that cross-building laser command injection is possible, at large distances and under realistic attack conditions. 
Finally, the experiment in Figure~\ref{fig:tower} was conducted at night due to safety requirements, with long-range attacks under illuminated conditions shown in Section~\ref{sect_corridor}.

\subsection{Attacking Authentication} \label{attacking_authentication}
Some of the current generation of VC systems attempt to protect unauthorized execution of sensitive commands by requiring additional user authentication step. For phone and tablet devices, the Siri and Alexa apps require the user to unlock the phone before executing certain commands (e.g., unlock front door, disable home alarm system). However, for devices that do not have other form of inputs beside the user's voice (e.g., voice-enabled smart speakers, cameras, and thermostats) a digit-based PIN code is used to authenticate the user before critical commands are performed.

\parhead{PIN Eavesdropping.}
The PIN number spoken by the user is inherently vulnerable to eavesdropping attacks, which can be performed remotely using a laser microphone (measuring the acoustic vibration of a glass window using a laser reflection~\cite{lasermic}), or using common audio eavesdropping techniques. Moreover, within an application the same PIN is used to authenticate more than one critical command (e.g., ``unlock the car'' and  ``start the engine'') while users often re-use PIN numbers across different applications. In both cases, increasing the number of PIN-protected commands ironically increases the opportunity for PIN eavesdropping attacks.

\parhead{PIN Brute forcing.} 
We also observed incorrect implementation of PIN verification mechanisms.
While Alexa naturally supports PIN authentication (limiting the user to three wrong attempts before requiring interaction with a phone application), Google Assistant delegates PIN authentication to third-party device vendors that often lack security experience.

Evaluating this design choice, we have investigated the feasibility of PIN brute forcing attacks on an August Smart Lock Pro, which is the most reviewed smart lock on Amazon at the time of writing. First, we have discovered that August does not enforce a reasonable PIN code length, allowing PINs containing anywhere from 1 to 6 digits for door unlocking. Next, we observed that August  does not limit the number of wrong attempts permitted by the user at the time of writing, nor does the lock implement a time delay mechanism between incorrect attempts, allowing the attacker o to unlock the target's door is to simply enumerating all possible PIN codes. 
 
 Empirically verifying this, we have written a program that enumerates all 4-digit PIN numbers using a synthetic voice. After each unsuccessful attempt, the Google home device responded with ``Sorry, the security code is incorrect, can I have your security code to unlock the front door?'' only to have our program speak the next PIN candidate. Overall, a single unlock attempt lasted about 13 seconds, requiring 36 hours to enumerate the entire 4-digit space (3.6 hours for 3 digits). In both the 3- and 4-digit case, the door was successfully unlocked when the correct PIN was reached.

\parhead{PIN Bypassing.}
Finally, we discovered that while commands like ``unlock front door'' for August locks or ``disable alarm system'' for Ring alarms require PIN numbers, other commands such as ``open the garage door'' using an assistant-enabled garage door opener\footnote{https://www.garadget.com/} often do not require any authentication. Thus, even if one command is unavailable, the attacker can often achieve similar goals by using other commands.

\subsection{Attacking Cars}
Many modern cars have Internet-over-cellular connectivity, allowing their owners to perform certain operations via a dedicated app on their mobile devices. In some cases, this connectivity has further evolved (either by the vendor or by a third-party) in having the target's car be connected to a VC system, allowing voice unlocking and/or pre-heating (which often requires engine start). Thus, a compromised VC system might be used by an attacker to gain access to the target's car. 

In this section we investigate the feasibility of such attacks, using two major car manufactures, namely Tesla and Ford. 

\parhead{Tesla.} Tesla cars allow their owner to interact with the car using a dedicated Tesla-provided phone app. After installing the app on our phone and linking it to a Tesla Model S vehicle, we installed the ``EV Car''\footnote{https://assistant.google.com/services/a/uid/000000196c7e079e?hl=en} integration, linking it to the vehicle. While ``EV Car'' is not officially provided by Tesla, after successful configuration using the vehicle's owner credentials, we were able to get several capabilities.  These included getting information about the vehicle's current location\footnote{Admittedly, the audible location is of little use to a remote attacker who is unable to listen in on the speaker's output.}, locking and unlocking the doors and trunk, starting and stopping the vehicle's charging and the climate control system. Next, we note that we were able to perform all of these tasks using only voice commands without the need of a PIN number or key proximity. Finally, 
we were not able to start the car without key proximity. 

\parhead{Ford Cars.} For newer vehicles, Ford provides a phone app called ``FordPass'', that connects to the car's Ford SYNC system, and allows the owner to interact with the car over the Internet. Taking the next step, Ford also provides a FordPass Google Assistant integration\footnote{https://assistant.google.com/services/a/uid/000000ac1d2afd15} with similar capabilities as the ``EV Car'' integration for Tesla. While Ford implemented PIN protection for critical voice commands like remote engine start and door unlocking, like in the case of August locks, there is no protection against PIN brute forcing. Finally, while we were able to remotely open the doors and start the engine, shifting the vehicle out of ``Park'' immediately stopped the engine, preventing the unlocked car from being driven.

\subsection{Exploring Stealthy Attacks}
The attacks described so far can be spotted by the user of the targeted VC system in three ways. First, the user might notice the light indicators on the target device following a successful command injection. Next, the user might hear the device acknowledging the injected command. Finally, the user might notice the spot while the attacker tries to aim the laser at the target microphone port.

While the first issue is a limitation of our attack (and in fact of any command injection attack), in this section we explore the attacker's options for addressing the remaining two issues.

\parhead{Acoustic Stealthiness.}
To tackle the issue of the device owner hearing the targeted device acknowledging the execution of voice command (or asking for a PIN number during the brute forcing process), the attacker can start the attack by asking the device to lower its speaker volume. For some devices (EcoBee, Google Nest Camera IQ, and Fire TV), the volume can be reduced to completely zero, while for other devices it can be set to barely-audible levels.
Moreover, the attacker can also abuse device features to achieve the same goal. For Google Assistant, enabling the ``do not disturb mode'' mutes reminders, broadcast messages and other spoken notifications. For Amazon Echo devices, enabling ``whisper mode'' significantly reduces the volume of the device responses during the attack to almost inaudible levels.

\parhead{Optical Stealthiness.} 
The attacker can also use an invisible laser wavelength to avoid having the owner spot the laser light aimed at the target device. However, as the laser spot is also invisible to the attacker, a camera sensitive to the appropriate wavelength is required for aiming.  
Experimentally verifying this, we replicated the attack on Google Home device from Section~\ref{sect_many_speakers} using a 980-nm infrared laser (Lilly Electronics 30 mW laser module).
We then connected the laser to a Thorlabs LDC205C driver, limiting its power to 5 mW. 
Finally, as the spot created by infrared lasers is invisible to humans, we aimed the laser using a smartphone camera (as these typically do not contain infrared filters). 

Using this setup, we have successfully injected voice commands to a Google Home at a distance of about 30 centimeters in the same setup as Section~\ref{sect_many_speakers}. The spot created by the infrared laser was barely visible using the phone camera, and completely invisible to the human eye. Finally, not wanting to risk prolonged exposure to invisible (but eye damaging) laser beams, we did not perform range experiments with this setup. However, given the color insensitivity described in Section~\ref{sec:feasibility}, we conjecture that results similar to those obtained in Section~\ref{sect_corridor} could be obtained here as well. 

\subsection{Avoiding the Need for Precise Aiming}\label{sec:flahlight}
Another limitation of the attacks described so far is the need to aim the laser spot precisely on the target's microphone ports. While we achieved such aiming in Section~\ref{sec:cross-building} by using geared camera tripod heads, in this section we show how the need for precise aiming can be avoided altogether.

An attacker can use a higher-power laser and trade its power with a larger laser spot size, which makes aiming considerably easier. Indeed, 
laser modules higher than 4W are commonly available on common e-commerce sites for laser engraving. Since we could not test such a high-power laser in an open-air environment due to safety concerns, we decided to use a laser-excited phosphor flashlight (Acebeam W30 with 500 lumens), which is technically a laser but sold as a flashlight with beam-expanding optics (making it a class 3B system).

To allow for voice modulation, we modified the flashlight by removing its original current driver and connecting its diode terminals to the Thorlabs LDC240C laser driver (see Figure~\ref{fig:flashlight}). Then, the experimental setup of Section~\ref{sect_corridor} is replicated except that the laser diode and telephoto lens is replaced with the flashlight. Using this setup, we successfully injected commands to a Google Home device at a range of about 10 meters, while running the flashlight at an output power of 1 W. Next, as can be seen in Figure~\ref{fig:flashlight}, the beam spot created by the flashlight is large enough to cover the entire target (and its microphone ports) without the need to use additional focusing optics and aiming equipment. However, we note that while the large spot size helps for imprecise aiming, the flashlight's quickly diverging beam also limits the attack's maximum range.

Finally, the large spot size  created by the flashlight (covering the entire device surface) can also be used to inject the sound  into  to multiple microphones simultaneously, thereby potentially defeating software-based anomaly detection countermeasures described in Section~\ref{sect_defense}.

\begin{figure}[t]
	\begin{center}
		\includegraphics[width=0.9\linewidth]{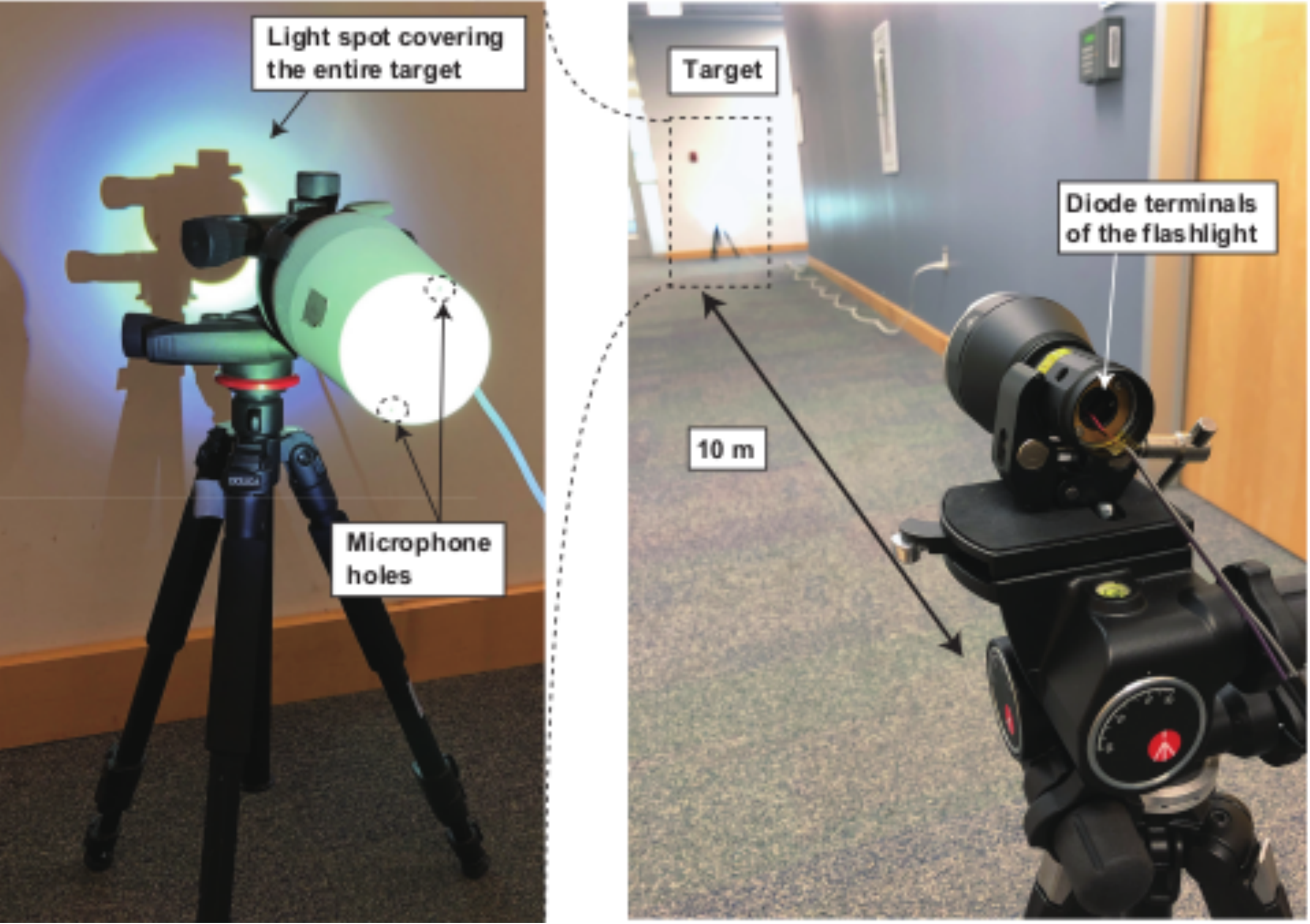}
	\end{center}
\vspace{-1.9em}
	\caption{Setup with laser flashlight to avoid precise aiming. (Left) Target device illuminated by the flashlight. (Right) Modified laser flashlight mounted on a geared tripod head aiming at the target 10 meters away.} \label{fig:flashlight}
\end{figure}

\subsection{Reducing the Attack Costs}
While the setups used for all the attacks described in this paper are built using readily available components, some equipment (such as the laser driver and diodes) are intended for lab use, making assembly and testing somewhat difficult for a non-experienced user. In this section we present a low-cost setup that can be easily constructed using improvised means and off-the-shelf consumer components.

\parhead{Laser Diode and Optics.} Modifying off-the-shelf laser pointers can be an easy way to get a laser source with collimation optics. In particular, cheap laser pointers often have no current regulators, having their anodes and cathodes directly connected to the batteries. Thus, we can easily connect a current driver to the pointer's battery connectors via alligator clips. Figure~\ref{fig:cheap_attack} shows a cheap laser pointer based setup, available at \$18 for 3 pieces at Amazon.

\begin{figure}[t]
	\begin{center}
\vspace{-1em}
		\includegraphics[width=0.99\linewidth]{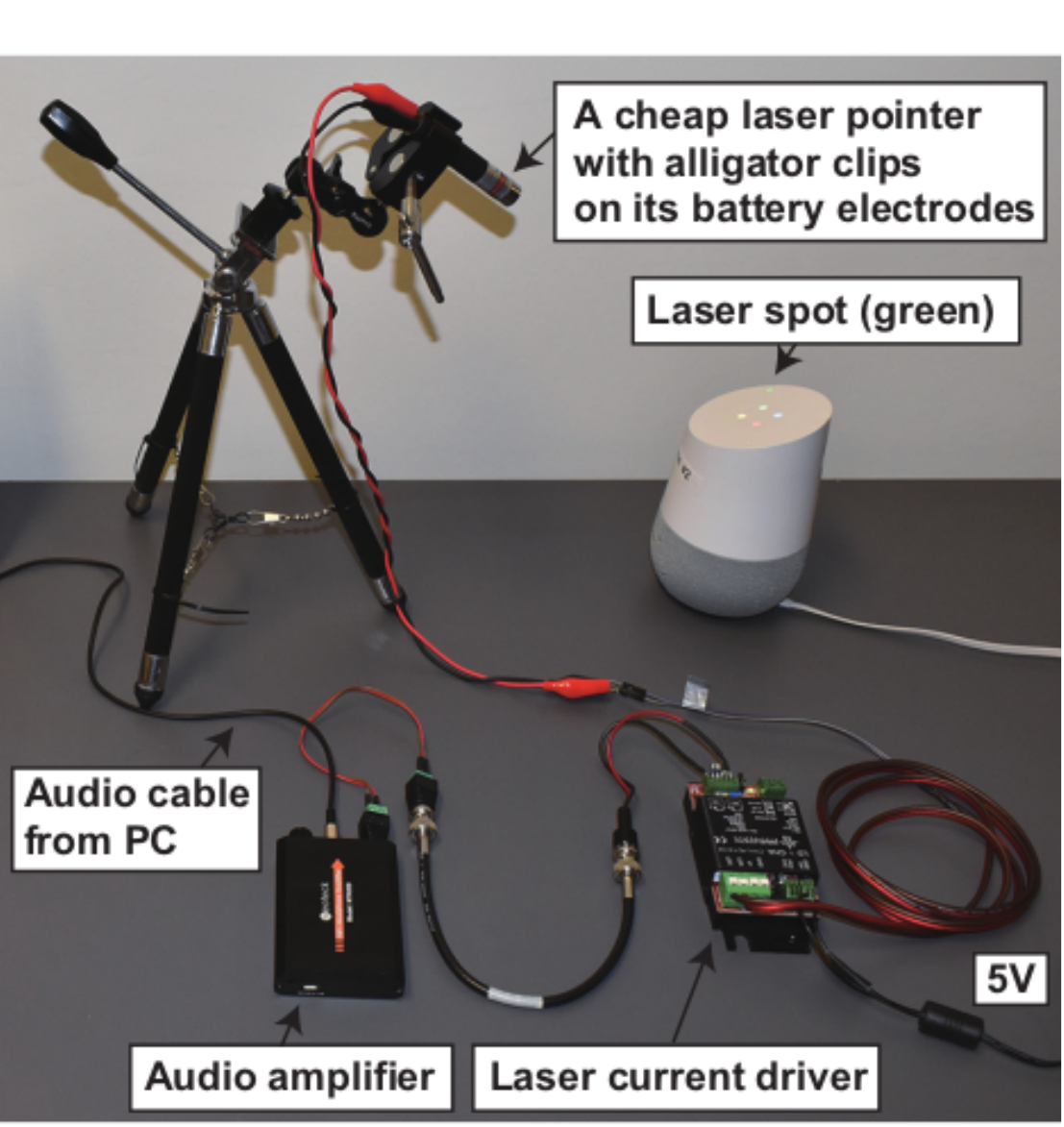}
	\end{center}
\vspace{-1.99em}
	\caption{Setup for low-cost attack: a laser current driver connected to a laser pointer attacking a Google Home device.} \label{fig:cheap_attack}
\end{figure}

\parhead{Laser Driver.}
The laser current driver with analog modulation port is the most specialized instrument of our setup, as we used the scientific-grade laser drivers that cost about \$1,500. However, cheaper alternatives exist, such as the Wavelength Electronics LD5CHA driver available for about \$300.

\parhead{Sound Source and Experimental Results.} Finally, the attacker needs a method for playing recorded audio commands. We used an ordinary on-board laptop sound card (Dell XPS 15 9570), amplified using a Neoteck NTK059 Headphone Amplifier (\$30 on Amazon).  See Figure~\ref{fig:cheap_attack} for a picture of a complete low-cost setup.  We have experimentally verified successful command injection using this setup into a Google Home located at a distance of 15 meters, with the main range limitation being the laser focusing optics and an artificially-limited power budget of 5 mW for safety reasons. Finally, we achieved a range of 110 meters with the cheap setup by replacing the laser optics with the telephoto lens from the previous sections.

\subsection{Attacking Non-MEMS Microphones}
Although  smart speakers, phones, and tablets typically use 
MEMS microphones due to their small footprint, we also investigate 
the feasibility of the attack on larger,
conventional non-MEMS microphones. We empirically verify this using a Sanwa 400-MC010 
 Electret Condenser Microphone, aiming the (blue) laser beam through the microphone's metallic mesh (See  Figure~\ref{fig:ecm} (Left)). Using the same parameters as in Section~\ref{sect_characterization} (e.g., $I_{DC}= 200$~mA and $I_{pp} = 150$~mA), we play a chirp signal varying  frequency linearly from 0 to 10 kHz in 5  seconds.  Figure~\ref{fig:ecm} (Right) shows the spectrogram of the audio recorded by the microphone, 
clearly showing repeated diagonal lines that correspond to the linear frequency sweep. We thus conclude that our results are also applicable beyond MEMS microphones, to electret condenser microphones.

\begin{figure}[t]
	\begin{center}
		\includegraphics[width=0.99\linewidth]{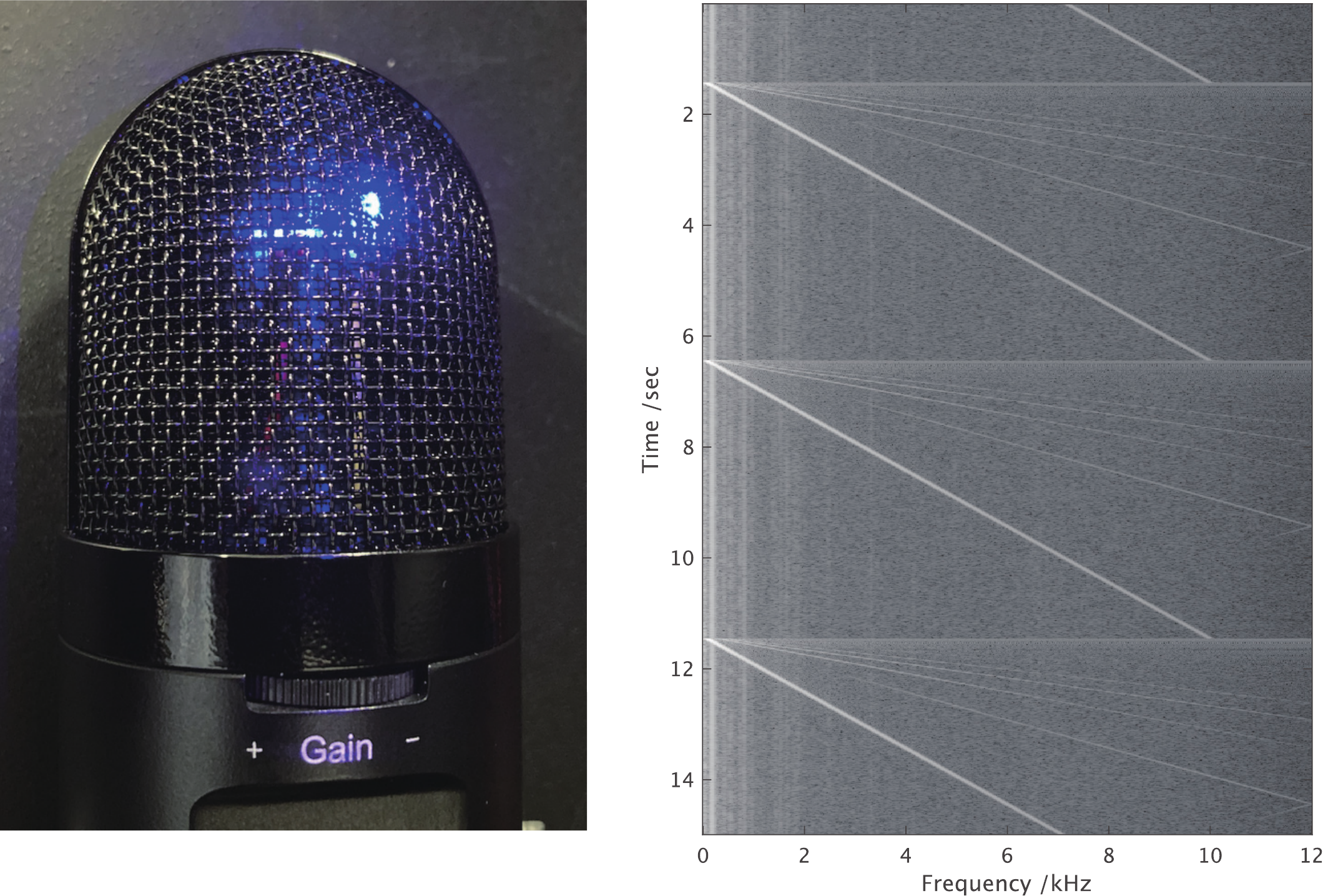}
	\end{center}
\vspace{-2.2em}
	\caption{(Left) Aiming a laser beam on an electret condenser microphone.
          (Right) Spectrogram of the microphone's output showing a clearly visible chirp signal.}
	\label{fig:ecm}
\end{figure}

\section{Countermeasures and Limitations}\label{sect_defense}

\subsection{Software-Based Approach}
As discussed in Section~\ref{attacking_authentication}, an additional layer of authentication can be effective at somewhat mitigating  the attack. Alternatively, in case the attacker cannot eavesdrop on the device's response (for example since the device is located far away behind a closed window), having the VC system ask the user a simple randomized question before command execution can be an effective way to prevent the attacker from obtaining successful command execution. However, note that adding an additional layer of interaction often comes at a cost of usability, limiting user adoption.

Next, manufacturers can attempt to use sensor fusion techniques~\cite{Davidson:2016:CUS:3027019.3027039} in the hopes of detecting light-based command injection. More specifically, voice assistants often have multiple microphones, which should receive similar signals due to the omnidirectional nature of sound propagation. Meanwhile, when the attacker uses a single laser, only one microphone receives a signal while the others receive nothing. Thus, manufacturers can attempt to mitigate the attack presented in this paper by comparing signals from multiple microphones, ignoring injected commands using a single laser. However, attackers can attempt to defeat such comparison countermeasures by simultaneously injecting light to all the device's microphones using multiple lasers or wide beams, see Section~\ref{sec:flahlight}. We leave this task of implementing such defenses and investigating their security properties to future work.

Finally, $\attack$ are very different compared to normal audible commands. For sensor-rich devices like phones and tablets, sensor-based intrusion detection techniques~\cite{sikder20176thsense} can potentially be used to identity and subsequently block such irregular command injection. We leave further exploration of this direction to future work. 

\subsection{Hardware-Based Approach} \label{sect_hw_defense}
It is possible to reduce the amount of light reaching the microphone's diaphragm using a barrier or diffracting film that physically blocks straight light beams,  while allowing sound waves to detour around it. Performing a literature review on proposed microphone designs, we have found several such suggestions, mainly aimed to protect microphones from sudden pressure spikes. 
For example, the designs in Figure~\ref{fig:mems_defense} have a silicon plate or movable shutter, both of which eliminate the line of sight to the diaphragm~\cite{7180939}. It is important to note however, that such barriers should be opaque to all light wavelengths (including infrared and ultraviolet), preventing the attacker from going through the barrier using a different colored light. Finally, a light-blocking barrier can be also implemented at the device level, by placing a non-transparent cover on top of the microphone hole, which attenuates the amount of light hitting the microphone.

\subsection{Limitations}

\parhead{Hardware Limitations.} Being a light-based attack, $\attack$ inherits all the limitations of light-related physics. In particular, $\attack$ assumes a line-of-sight threat model and does not properly penetrate opaque obstacles which might be penetrable to sound. Thus, even if attacking fabric-covered devices is sometimes possible (Section~\ref{sect_corridor}, Google Home Mini), we believe that for fabric-covered microphones' ports, the thickness of the cover can prevent successful attacks (e.g., in the case of Apple Homepods). We leave the analysis of such scenarios to future work.

In addition, unlike sound, $\attack$ requires careful aiming and line of sight access. In our experiments, we show how to partially overcome this limitation by using a telescope to remotely determine the assistant type and location of the microphones from the device's appearance.

Finally, while line of sight access is often available for smart speakers visible through windows, the situation is different for mobile devices such as smart watches, phones and tablets. This is since unlike static smart speakers, these devices are often mobile, requiring an attacker to quickly aim and inject commands. When combined with the precise aiming and higher laser power required to attack such devices, successful  $\attack$ attacks might be particularly challenging. We thus leave the task of systematically exploring such devices to future work. 

\parhead{Liveness Test and Continuous Authentication.}Unlike some other injection attacks, $\attack$' threat model and lack of proper feedback channels make it difficult for the attacker to pass any sorts of liveness checks or continuous authentication methods. These can be as primitive as asking a user simple questions before performing a command, or as sophisticated as using data from different microphones~\cite{DBLP:conf/ccs/ZhangT0016, DBLP:conf/ccs/ZhangT017,lu2018lippass},  sound reflections~\cite{lu2019lip}, or other sensors~\cite{feng2017continuous}  to verify that the incoming commands were indeed spoken by a live human. We leave the task of implementing such defenses in deployed VC systems as an avenue for future works.

\begin{figure}[t]
	\begin{center}
		\includegraphics[width=0.9\linewidth]{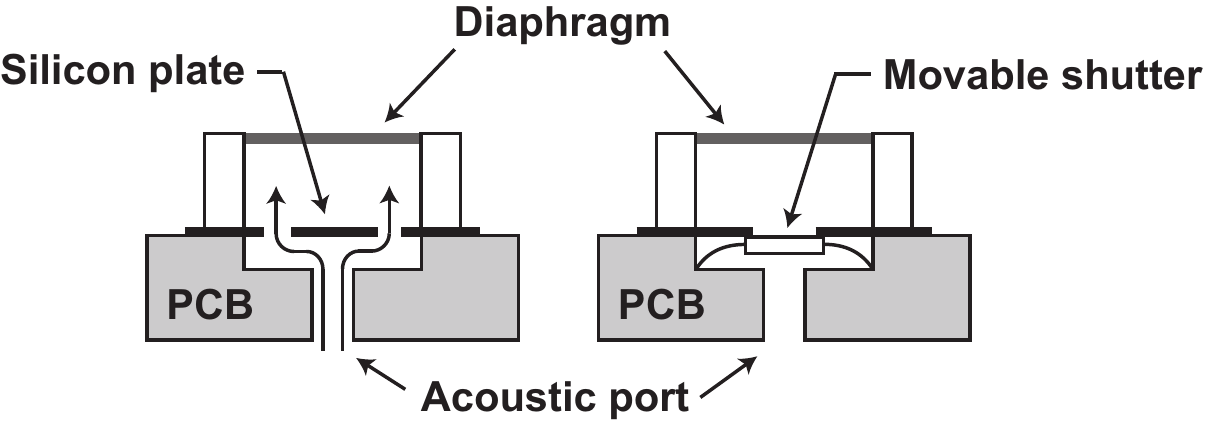}
	\end{center}
\vspace{-1.99em}
	\caption{Designs of MEMS microphone with light-blocking barriers~\cite{7180939}}
	\label{fig:mems_defense}
\end{figure}

\section{Conclusions and Future Work} \label{conclusion}

In this paper we presented $\attack$, which is an attack that uses light to inject commands into voice-controllable systems from large distances. To mount the attack, we transmit light modulated with an audio signal, which is converted back to audio within the microphone. We demonstrated $\attack$ on many commercially-available voice-controllable systems that use Siri, Portal, Google Assistant, and Alexa, obtaining successful command  injections at a distance of more than 100 meters while  penetrating clear glass windows. Next, we highlight deficiencies in the security of voice-controllable systems, which leads to additional compromises of third-party hardware such as locks and cars.

Better understanding of the physics behind the attack will benefit both new attacks and countermeasures.
In particular, we can possibly use the same principle to mount other acoustic injection attacks (e.g., on motion sensors) using light. In addition, heating by laser can also be an effective way of injecting false signals to sensors. 

\section{Acknowledgments}
We thank John Nees for advice on laser operation and laser optics.  This research was funded by JSPS KAKENHI Grant \#JP18K18047 and \#JP18KK0312, by
DARPA and AFRL under contracts
FA8750-19-C-0531 
and HR001120C0087, 
by NSF under grants CNS-1954712 
and CNS-2031077, gifts from Intel, AMD, and Analog Devices, and an award from MCity at the University of Michigan. 

\bibliographystyle{IEEEtranN}
\bibliography{laser}

\end{document}